%% file: sample-manuscript.tex
\documentclass[screen]{acmart}
\setcopyright{none}
\AtBeginDocument{%
  }

\setcopyright{acmlicensed}
\copyrightyear{2026}
\acmYear{2026}
\acmDOI{10.1145/3797029}
\usepackage{multirow}
\usepackage[nolist]{acronym}
\usepackage{pifont}
\usepackage{graphicx}
\usepackage[inkscapelatex=false]{svg}
\usepackage{algorithm}
\usepackage{algpseudocode}
\usepackage{amsmath}
\usepackage{tikz}
\usepackage{xcolor}

\begin{document}
\include{./acro}
\title{EgoAdapt: Enhancing Robustness in Egocentric Interactive Speaker Detection Under Missing Modalities}
\thanks{This work is supported in part by the National Natural Science Foundation of China under grants 62272229 and 62306029, Shenzhen Science and Technology Program JCYJ20230807142001004, Beijing Natural Science Foundation (L233032) and Young Elite Scientists Sponsorship Program of the Beijing High Innovation Plan.}

\author{Xinyuan Qian}
\authornote{Both authors contributed equally to this research.}
\affiliation{%
  \institution{School of Computer and Communication Enngineering, University of Science and Technology Beijing}
  \city{Beijing}
  \country{China}
}
\email{qianxy@ustb.edu.cn}

\author{Xinjia Zhu}
\authornotemark[1]
\affiliation{%
  \institution{MIIT Key Laboratory of Pattern Analysis and Machine Intelligence, College of Computer Science and Technology, Nanjing University of Aeronautics and Astronautics}
  \city{Nanjing}
  \country{China}
}
\email{xinjiazhu@nuaa.edu.cn}

\author{Alessio Brutti}
\affiliation{%
  \institution{Fondazione Bruno Kessler}
  \city{Trento}
  \country{Italy}
}
\email{brutti@fbk.eu}

\author{Dong Liang}
\authornote{Corresponding author.}
\affiliation{%
  \institution{MIIT Key Laboratory of Pattern Analysis and Machine Intelligence, College of Computer Science and Technology, Shenzhen Research Institute, Nanjing University of Aeronautics and Astronautics}
  \city{Nanjing}
  \country{China}
}

\email{liangdong@nuaa.edu.cn}

\renewcommand{\shortauthors}{Qian et al.}

\begin{abstract}
\ac{TTM} task is a pivotal component in understanding human social interactions, aiming to determine who is engaged in conversation with the camera-wearer.
% identify the interactive speaker addressing the camera wearer from an egocentric perspective. 
Traditional models often face challenges in real-world scenarios due to missing visual data, neglecting the role of head orientation, and background noise. 
This study addresses these limitations by introducing \textbf{EgoAdapt}, an adaptive framework designed for robust egocentric `Talking to Me' speaker detection under missing modalities.
Specifically, EgoAdapt incorporates three key modules: 
(1) a \textbf{V}isual \textbf{S}peaker \textbf{T}arget \textbf{R}ecognition (VSTR) module that captures head orientation as a non-verbal cue and lip movement as a verbal cue, allowing a comprehensive interpretation of both verbal and non-verbal signals to address \ac{TTM}, setting it apart from tasks focused solely on detecting speaking status;
(2) a \textbf{P}arallel \textbf{S}hared-weight \textbf{A}udio (PSA) encoder for enhanced audio feature extraction in noisy environments, and 
(3) a \textbf{V}isual \textbf{M}odality \textbf{M}issing \textbf{A}wareness (VMMA) module that estimates the presence or absence of each modality at each frame to adjust the system response dynamically.
Comprehensive evaluations on the TTM benchmark of the Ego4D dataset demonstrate that EgoAdapt achieves a 
\ac{mAP} of 67.39\% and an \ac{Acc} of 62.01\%, significantly outperforming state-of-the-art method by 4.96\% in \ac{Acc} and 1.56\% in \ac{mAP}.
\end{abstract}




\maketitle

\section{INTRODUCTION}
Understanding social interactions is crucial for embodied agents in analyzing the intricate mechanisms behind human communications. 
Traditional social interaction methods mainly use speech signals, where advances in speech separation and extraction~\cite{overview2018,qian2025sav,jiang2025tpeech,luo2018tasnet}, \ac{ASR} ~\cite{yu2016automatic,wang2024predict,malik2021automatic,benzeghiba2007automatic,kim2022squeezeformer}, and dialog management ~\cite{bordes2016learning,harms2018approaches,bohus2009ravenclaw} have improved the model's ability to process and understand verbal communications.
However, these auditory methods encounter considerable challenges, especially in complex multi-speaker scenarios.
For example, speech separation approaches often struggle to robustly determine the number of speakers and attribute speech to the correct person ~\cite{wang2018supervised,luo2019conv,subakan2021attention}. When people speak simultaneously, it is also difficult for \ac{ASR} models to recognize the spoken content of each individual ~\cite{strik1999modeling,larson2012spoken,lee2015spoken}.
Thus, it is important to introduce a front-end module to determine when, what, and to whom to respond.
In contrast to auditory methods, vision-based methods~\cite{voulodimos2018deep, szeliski2022computer, guo2022attention} are immune to background noise or overlapping speech, offering a distinct advantage in environments where auditory signals are obscured. 
In particular, visual cues are beneficial when multiple conversations co-occur or in difficult noise situations, which can providing clarity and focus on non-verbal aspects of communication, such as facial expressions ~\cite{calder2016understanding,li2020deep,ge2022facial,adyapady2023comprehensive},  gaze estimation~\cite{cheng2024appearance,kar2017review,wang2002study,wang2021vision,zhang2015appearance}, head 
pose estimation~\cite{ruiz2018fine,ranjan2017hyperface,murphy2008head,meyer2015robust,fanelli2011real}, and body language~\cite{beck2012emotional, martinez2016contributions,rastgoo2024survey, qian2025dual}.
Despite these advances, visual methods also have limitations, particularly in distinguishing between relevant social cues and other facial or mouth movements, such as eating, yawning, or making emotional expressions. 
These unrelated motions can be mistakenly interpreted as meaningful social signals, which eventually affect the accuracy of visual interaction interpretation.

\begin{figure}[ht]
    \centering
    \includegraphics[width=0.95\textwidth]{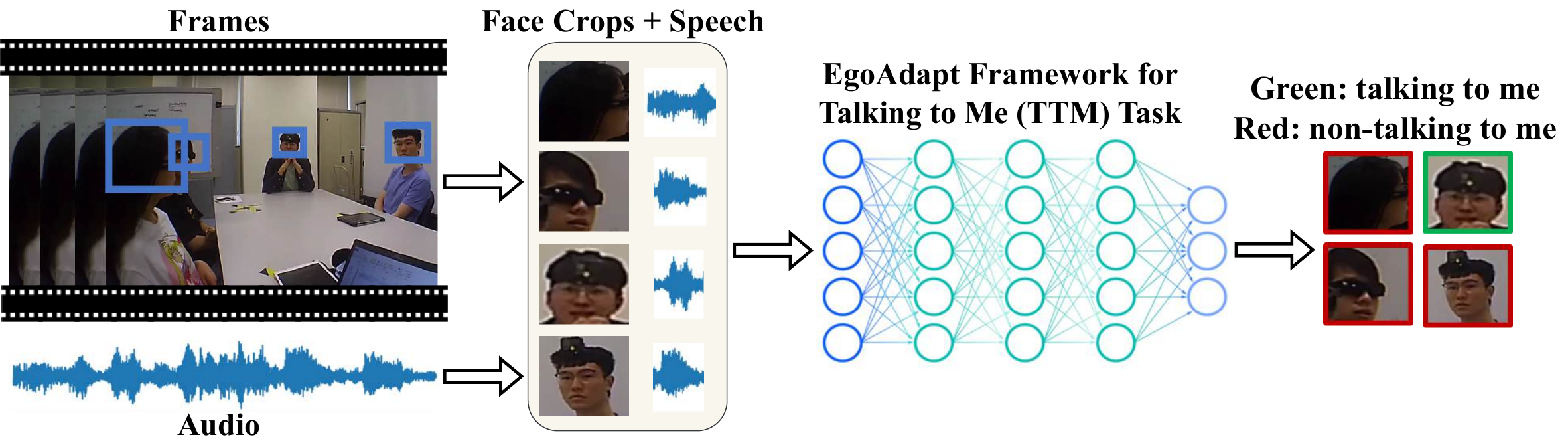} 
    \caption*{(a) The \ac{TTM} Task Setting }
    \vspace{0.2cm} 
    \begin{minipage}[t]{0.31\textwidth}
        \centering
        \includegraphics[width=\textwidth]{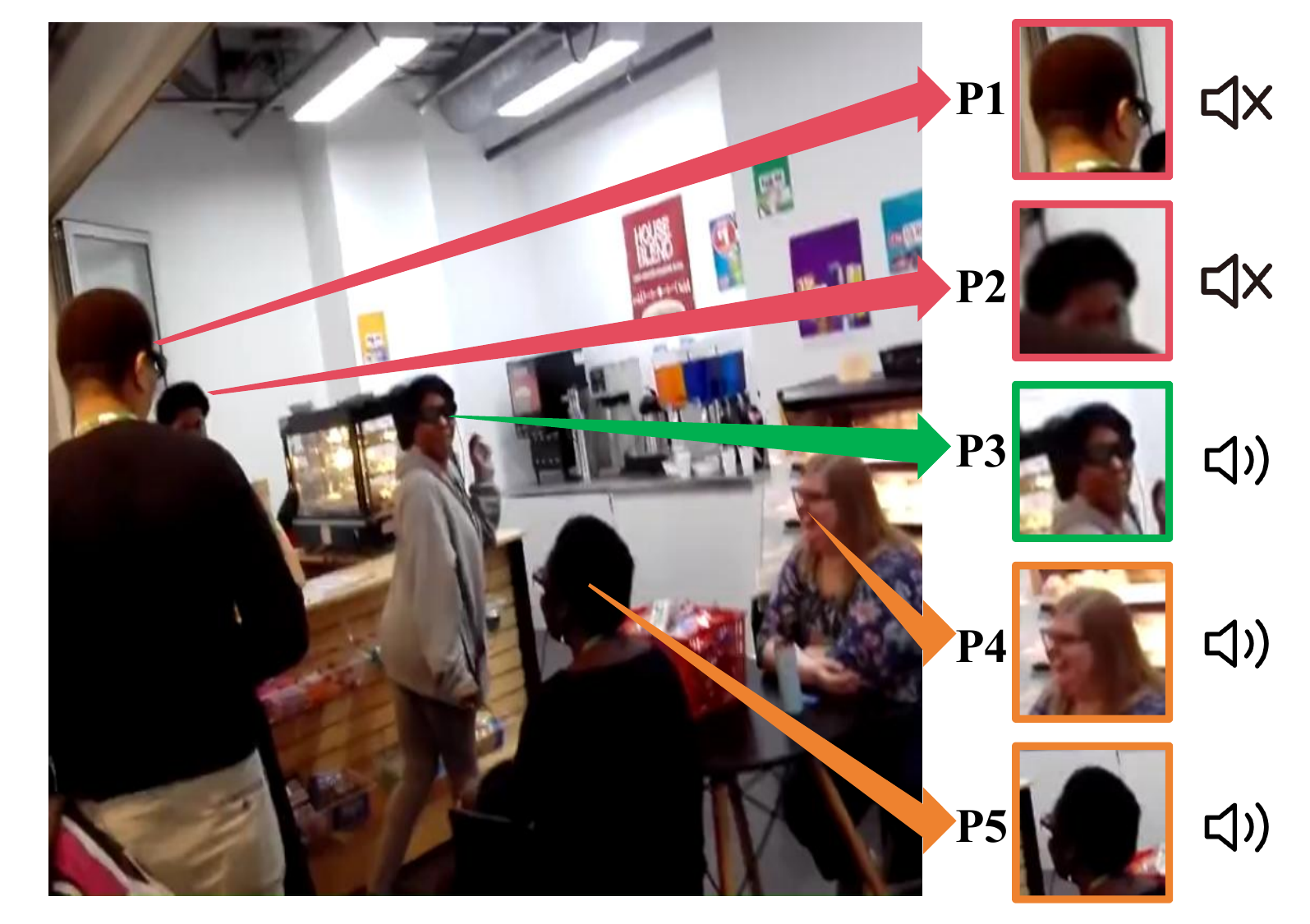}
        \caption*{(b) Multi-Speaker Scenario }
    \end{minipage}
    \hfill
    \begin{tikzpicture}
        \draw[dashed] (0,0) -- (0,3); 
    \end{tikzpicture}
    \hfill
    \begin{minipage}[t]{0.31\textwidth}
        \centering
        \includegraphics[width=\textwidth]{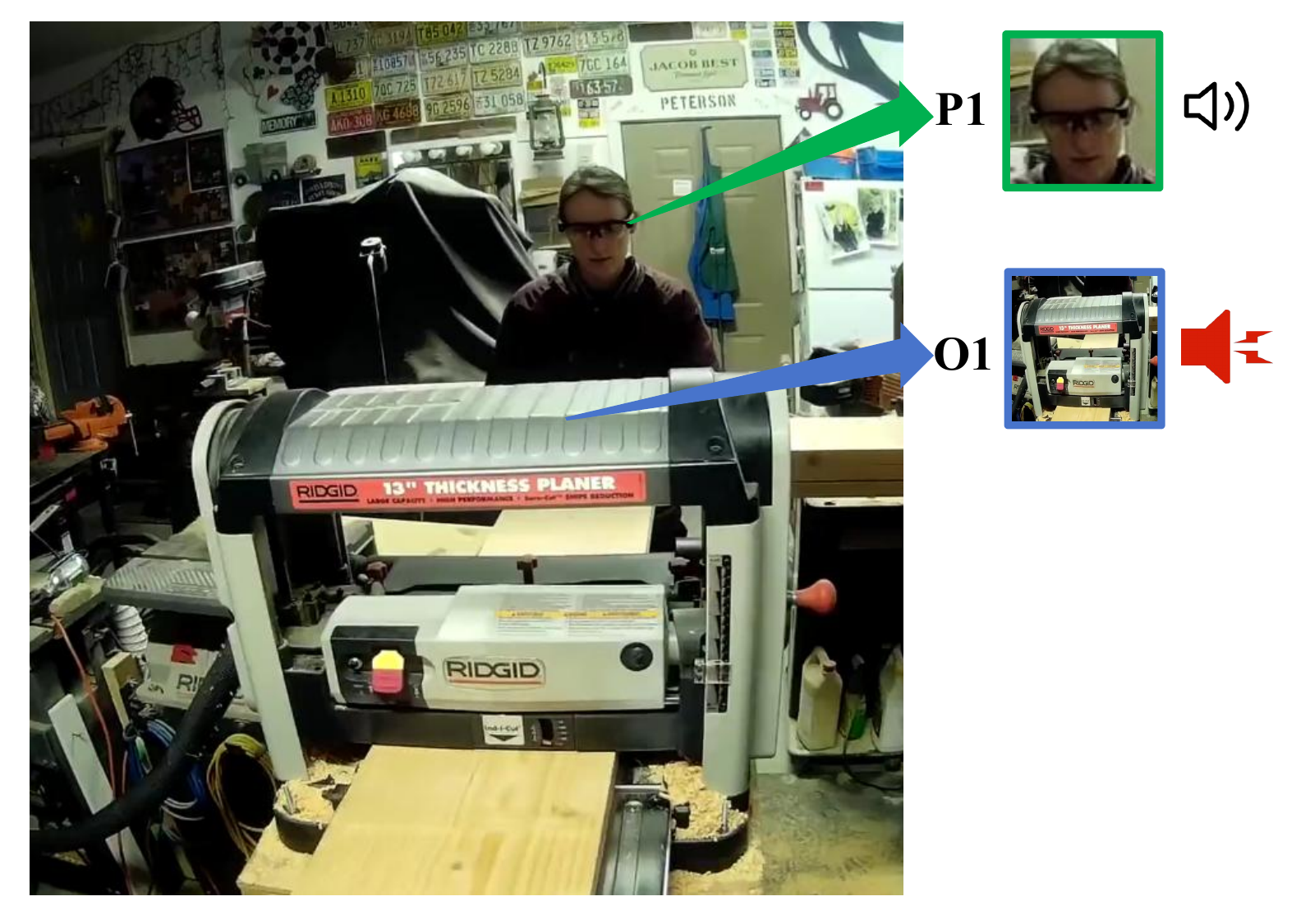}
        \caption*{(c) Background Noise Scenario}
    \end{minipage}
    \hfill
    \begin{tikzpicture}
        \draw[dashed] (0,0) -- (0,3); 
    \end{tikzpicture}
    \hfill
    \begin{minipage}[t]{0.31\textwidth}
        \centering
        \includegraphics[width=\textwidth]{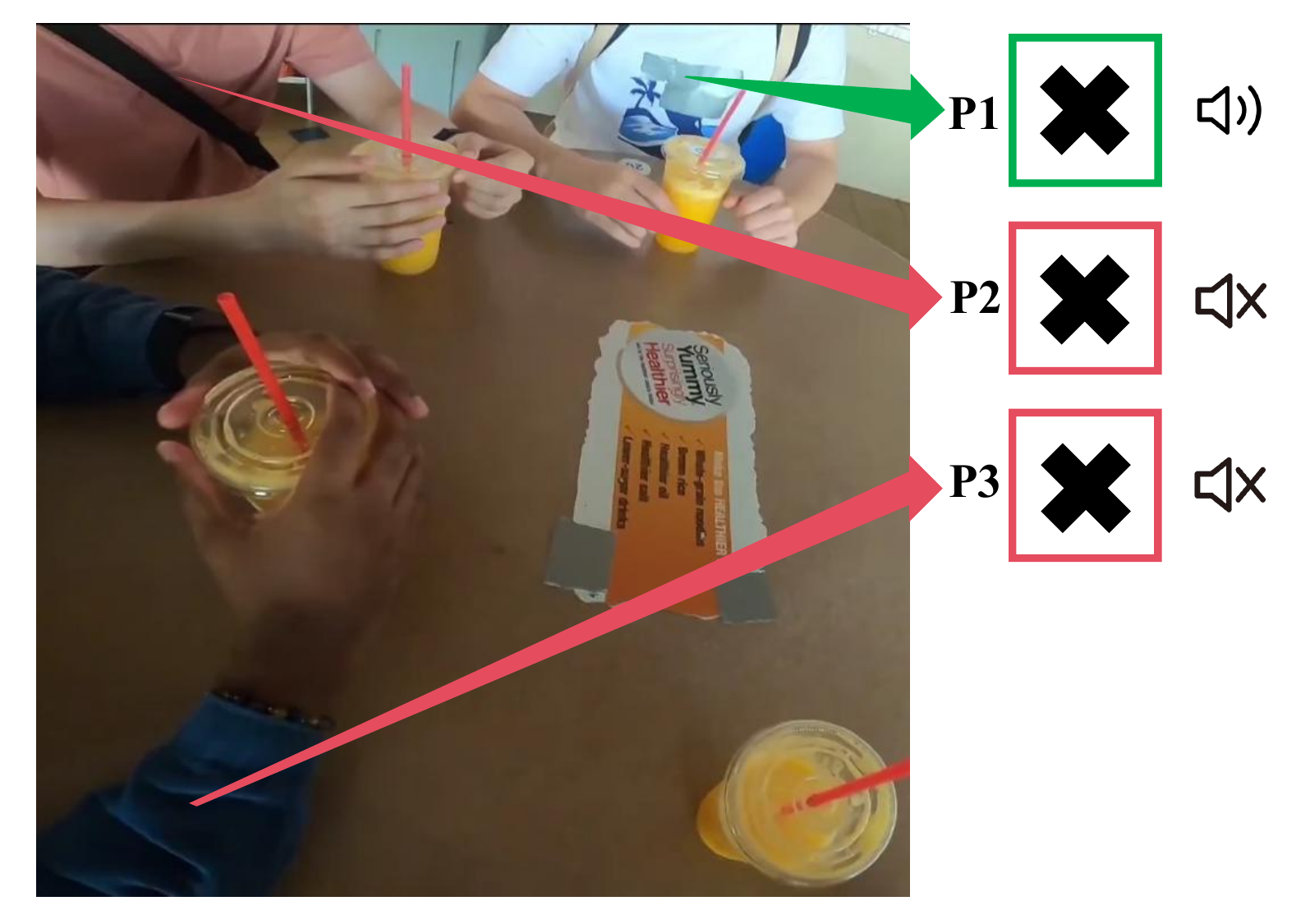}
        \caption*{(d) Occluded Heads Scenario}
    \end{minipage}
    \caption{
        \ac{TTM} task setting and challenges.
        (a) The \ac{TTM} task consists in the frame-level identification of which person is talking to the camera wearer, given the synchronized audio-visual streams from the first-person perspective. (b)-(d) An overview of various real-life scenarios depicted in the egocentric video. (The red bounding box denotes a non-speaking person; the green bounding box denotes a person who is talking to the camera wearer; the blue bounding box denotes a sound source other than humans; the orange bounding box denotes a person who is talking but not talking to the camera wearer; the box with a cross indicates that no face was detected).
        }
    \label{fig:task and challenges}
\end{figure}

There has been a growing interest in combining audio-visual strengths to overcome the unimodal limitations to gain a more comprehensive understanding of social interactions.
The \ac{TTM} task, introduced along with the Ego4D dataset ~\cite{grauman2022ego4d}, determines which person is talking to the camera wearer given the synchronized audio-visual streams, as illustrated in Fig.~\ref{fig:task and challenges}~(a). 
While previous \ac{TTM} methods ~\cite{grauman2022ego4d, xue2023egocentric, lin2023quavf} mainly focus on the exploration of different audio-visual fusion strategies, we particularly tackle realistic challenges, as illustrated in Fig.~\ref{fig:task and challenges}~(b)-(d).
We consider the following realistic challenges:
(1) Multi-Speaker Scenario:
It is crucial to initially evaluate the individual's speech initiation and subsequently assess the speaker's head orientation to determine if they are talking to the camera wearer. The integration of audio and visual cues is pivotal for this assessment.
(2) Background Noise Scenario: 
Existing methods make it difficult to distinguish speakers under intense background noise, which may lead to false or missed detection of the speaker.
(3) Occluded Heads Scenario: 
In situations where the camera's movement restricts the field of view, identifying the speaker becomes more challenging. 
Consequently, reliance on auditory cues becomes essential. 
Without visual indicators, audio signals can be analyzed to determine if someone is speaking toward the camera wearer. 

To address the aforementioned challenges, we propose EgoAdapt, a novel method to enhance robustness in egocentric interactive speaker detection under missing modalities. To sum up, our contributions can be summarized as follows:

\begin{itemize}
\item  \ac{VSTR} module for complex social cues: To better interpret speaker intent, the \ac{VSTR} module integrates head orientation (non-verbal cue) and lip movement (verbal cue) as essential social signals. This approach allows EgoAdapt to more accurately determine when a speaker is directly addressing the camera wearer, making it suitable for complex social interactions rather than simple speaking-status detection.
\item  \ac{PSA} encoder for noise-resilient audio processing: EgoAdapt includes a specialized \ac{PSA} encoder, which enhances the model’s ability to extract salient audio features even under substantial background noise.
\item  \ac{VMMA} module for dynamic modality adaptation: We introduce a \ac{VMMA} module, which dynamically assesses the presence or absence of head information in each frame. This module enables EgoAdapt to adjust its reliance on audio or visual cues based on real-time data availability, ensuring robust performance even in scenarios with occluded or absent visual head data.
\item Experimental results show that our method achieves competitive performance on the \ac{TTM} benchmark of the Ego4D dataset, surpassing existing baseline methods. 
\end{itemize}

\section{RELATED WORK}
Egocentric video understanding is essential for applications such as human–robot interaction and assistive technologies, yet it faces unique challenges from camera motion, occlusion, and dynamic social settings.  
The TTM task aims to determine whether a visible participant addresses the camera wearer by leveraging multimodal cues from head pose, lip motion, and audio.  
Real-world conditions often introduce missing or degraded modalities, such as noisy speech, visual occlusion, or audio–visual misalignment.  
These challenges demand robust multimodal fusion methods capable of adapting to incomplete or corrupted inputs.  
This motivates the exploration of technologies that integrate discriminative features across modalities while ensuring resilience to modality loss.  

\subsection{Audio-Visual Active Speaker Detection}
Audio-visual active speaker detection is a technique for identifying speakers that appear in synchronized audio-visual streams~\cite{10814693,9858007,robi2024active}. It determines the position and identity of speakers by analyzing the sound characteristics in audio or video signals, such as spectrum, sound intensity, and time-domain characteristics. The detection of audio and video speakers is commonly used in applications such as speech recognition~\cite{li2022recent, wang2019overview,liu2025machine,ahlawat2025automatic}, audio processing~\cite{luo2019fasnet, zolzer2022digital, mcloughlin2009applied}, and video surveillance~\cite{elharrouss2021review, haering2008evolution, tsakanikas2018video}, which can help the system identify and separate different sources of sound with greater accuracy. The extended UniCon ~\cite{zhang2021ictcas} introduced a spatial context to obtain a robust model by integrating three types of contextual information. In contrast, TalkNet ~\cite{tao2021someone} achieved better performance using cross-attention and self-attention modules to aggregate audio and visual features. Subsequently, based on this work, ASD-Transformer~\cite{datta2022asd} improved the performance by introducing positional encoding and improving the attention module. To better exploit the potential of the attention module, ADENet ~\cite{xiong2022look} introduced multi-modal layer normalization to alleviate the distribution misalignment of audio-visual features.
Despite these advances, existing audio-visual active speaker detection methods typically assume that both modalities are complete and of high quality, limiting their robustness in realistic egocentric scenarios where audio can be corrupted by environmental noise, visual cues (e.g., lip motion) may be occluded, and cross-modal synchronization may be imperfect. Moreover, most prior approaches focus on generic speaker detection and do not explicitly address the “Talking-to-Me” addressee recognition problem, which requires fine-grained modeling of interactions between the camera wearer and surrounding speakers.
\subsection{Audio-Visual Social Interaction}
Audio-visual social interaction refers to the simultaneous involvement of multiple different communication modes in social interaction, such as language, facial expressions, gestures, actions, etc. This multimodal social interaction can provide richer and more natural information for human-computer interaction and machine learning~\cite{friedland2010dialocalization,biel2011vlogsense,10184468}. The \ac{TTM} task is a specific application scenario proposed in the Ego4D dataset ~\cite{grauman2022ego4d}, which involves identifying whether someone is talking to the camera wearer in a video. It requires the model to not only analyze visual information in the video (such as head posture, facial expressions) but also combine auditory information (such as speaking voice, speech features). Ego4D baseline for \ac{TTM}~\cite{grauman2022ego4d} introduces a method that utilizes ResNet-18 ~\cite{he2016deep} for extracting facial features and ResNet-SE~\cite{hu2018squeeze} for obtaining audio embeddings. The audio and video embeddings are subsequently concatenated and fed into a fully connected layer to predict the \ac{TTM} task. Subsequently, based on this work, EgoT2~\cite{xue2023egocentric} refines the outputs from multiple models, each optimized for distinct tasks. QuAVF ~\cite{lin2023quavf} proposes two separate models to process input videos and audio. In addition, it uses a facial quality score derived from a facial landmark prediction model to filter noisy face input data and integrates results from the two branches. 
Ex2Eg-MAE~\cite{10.1007/978-3-031-72989-8_1} proposes a novel framework leveraging novel-view face synthesis for dynamic perspective data augmentation from abundant exocentric videos and enhances self-supervised learning process for VideoMAE~\cite{tong2022videomae}.

Unlike traditional speaker detection tasks that operate in static or controlled environments, egocentric detection deals with first-person, dynamic viewpoints, often involving abrupt camera movements, partial occlusions, and challenging background noise. 
Previous egocentric interactive speaker detection methods have primarily focused on audio-visual active speaker detection without delving into the effectiveness of the interaction between verbal and non-verbal cues for the interaction with the camera wearer. 
Moreover, the diverse nature of egocentric interaction scenarios often involves background noise interference in real-world settings. 
Furthermore, although some studies ~\cite{grauman2022ego4d,lin2023quavf} have acknowledged the issue of missing visual data, they have not addressed how to improve the model robustness against modality loss.
Recognizing these limitations, our proposed EgoAdapt framework introduces a novel \ac{VMMA} module that dynamically adjusts system responses based on the availability of visual and audio input, allowing it to handle  missing visual data in real time. 
Furthermore, we incorporate a \ac{VSTR} module that explicitly interprets head orientation as a non-verbal cue and lip movement as a verbal cue to improve the model's capability in distinguishing between multiple speakers in egocentric settings. 
As for body-part cues (e.g., hands and torso), while they may carry potentially rich information, they are often difficult to obtain reliably under the egocentric setting. Frequent occlusions, incomplete visibility, and motion blur from the moving camera severely undermine their consistency. In contrast, head and speech modalities remain relatively stable and controllable. For this reason, our current framework focuses on these more reliable signals to ensure robustness and reproducibility.
By combining these modules with a \ac{PSA} encoder designed to reduce noise interference, EgoAdapt offers a robust solution for the TTM task, effectively addressing the limitations of current audio-visual speaker detection approaches.

\section{METHODOLOGY}

\begin{figure*}[t]
\centering
\includegraphics[width=0.95\textwidth]{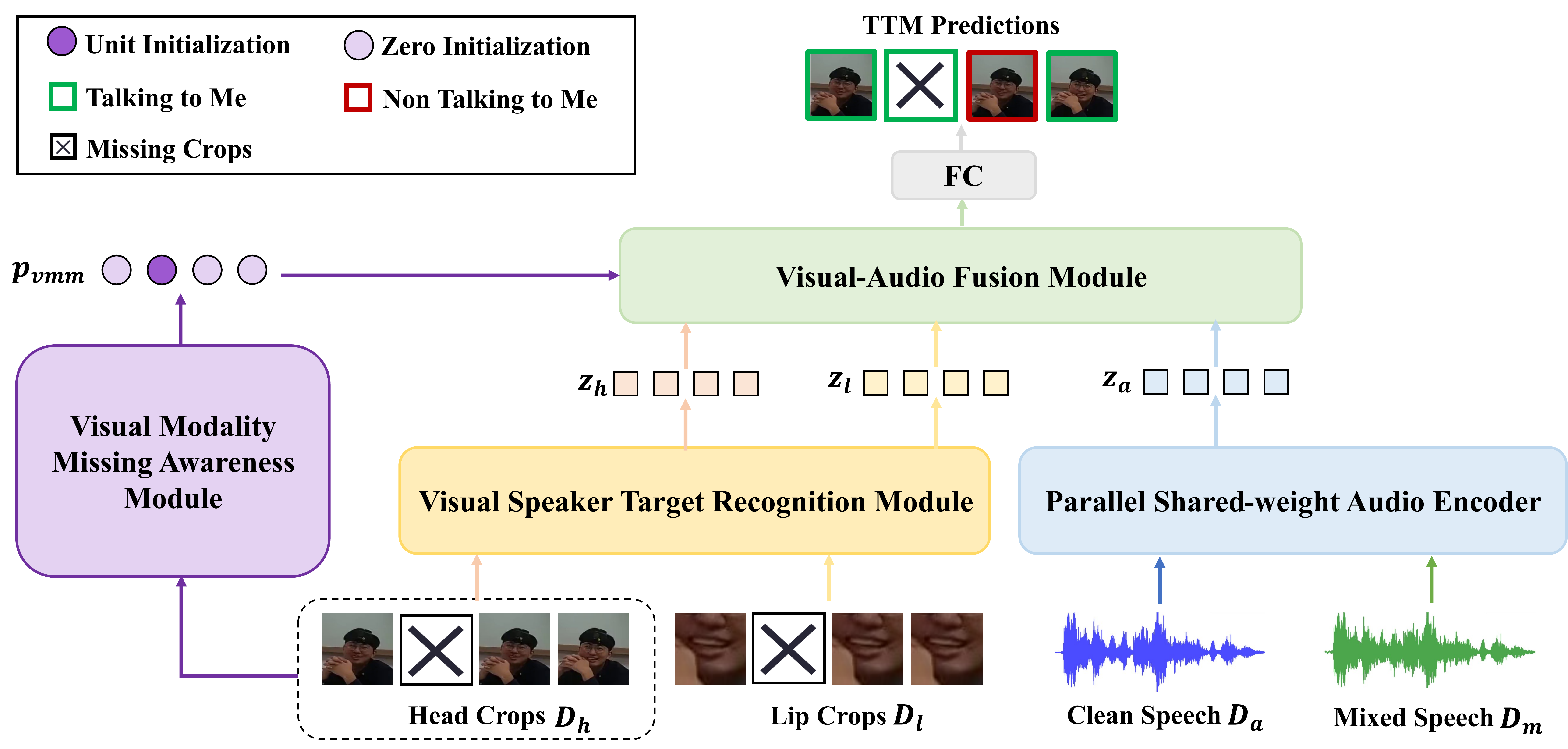} 
\caption{
Overview of the \textbf{EgoAdapt} framework for egocentric interactive speaker detection. 
The \acf{VMMA} module assesses modality availability and outputs \( p_{vmm} \) to guide adaptation. 
The \acf{VSTR} module  extract non-verbal and verbal cues (\( z_h \) and \( z_l \)) from head crops \( D_h \) and lip crops \( D_l \), respectively, while the \acf{PSA} encoder processes clean and noisy audio (\( D_a \) and \( D_m \)) to produce \( z_a \). 
These cues are fused in the Visual-Audio Fusion module for the final TTM Prediction of "Talking to Me" or "Non Talking to Me".
 }
\label{fig:EgoAdapt}
\end{figure*}

\subsection{Problem Definition and Method Overview}
Given a synchronized audio-visual stream, we aim to determine whether a participant in a video frame is talking to the camera wearer or not. 
Specifically, we define head crops \( D_h = \{h_t\}_{t=1}^{T} \) as the head of the detected speaker in each frame where $t=1,...,T$ with \( T \)  the total frame number and lip crops \( D_l = \{l_t\}_{t=1}^{T} \) as the extracted lip region.
The corresponding audio segments are denoted as \( D_a = \{a_t\}_{t=1}^{T} \)
and the \ac{TTM} problem can be formulated as:
\begin{equation}
    \hat{y}_t = \mathcal{F}(D_h, D_l, D_a| \Omega)
\end{equation}
where $\mathcal{F}(\cdot)$ denotes the neural network with trainable parameters $\Omega$. $\hat{y}_t$ is the estimate of \ac{TTM} label ${y}_t$ that is set to 1 if the participant in frame \( t \) is talking to the camera wearer and 0 otherwise.

Our proposed EgoAdapt framework is shown in Fig.~\ref{fig:EgoAdapt}. 
In particular, our framework addresses three key challenges in egocentric audio-visual data processing: 

(Q1) How can visual information be used to learn non-verbal cues for accurately identifying the person talking to the camera wearer? 
To determine if someone is talking to the camera wearer, we use the \ac{VSTR} module to extract head orientation and lip features, capturing both non-verbal and verbal cues from video frames (Section 3.2).

(Q2) How can the \ac{TTM} task be addressed in noisy environments? 
To maintain the system robustness, the proposed \ac{PSA} encoder processes both clean and mixed audio segments using shared weights to handle variable noise levels, with the network trained on \ac{MSE} loss (Section 3.3).

(Q3) How can the framework cope with situations where the visual modality is sometimes missing? 
To tackle incomplete visual information, the \ac{VMMA} module dynamically evaluates the availability of both visual and audio cues. It outputs \( p_{vmm} \), which directs the model in adjusting to modality variations, ensuring stable performance even in the absence of visual cues (Section 3.4).

The Visual-Audio Fusion module then integrates features from all modules, allowing the system to produce accurate frame-level predictions of whether the person is "Talking to Me" even under severe acoustic conditions (Section 3.5).

\subsection{Visual Speaker Target Recognition Module}

\begin{figure*}[ht]
\centering
\includegraphics[width=0.4\textwidth]{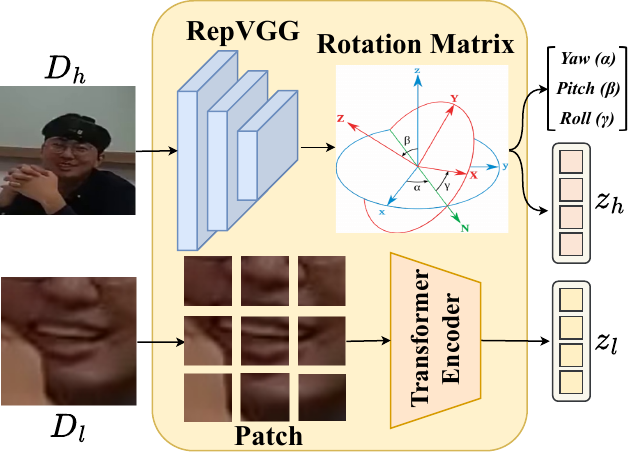} 
\caption{ \acf{VSTR} module. It includes two components: the Head Pose Feature Extraction Module, which captures head orientation (non-verbal cues), and the Lip Feature Extraction Module, which extracts lip movement features (verbal cues). } 

\label{fig:VSTR}
\end{figure*}

Our vision branch is designed to extract and analyze critical information about the speaker's participation.
When an active speaker is detected, it is imperative to determine the speaking target,  which identifies the individual or entity to whom the speaker is directing their speech. In typical interactive scenarios, as illustrated in Fig.~\ref{fig:task and challenges}~(b), there is a notable tendency to face-to-face orientation when the camera wearer is conversing with a speaker. Therefore, 
\ac{VSTR} becomes particularly relevant to discover whether the speaker is talking to the camera wearer directly.
As illustrated in Fig.~\ref{fig:VSTR}, to simultaneously capture both the visual speaking features, e.g., lip motions and head orientations, our approach integrates two distinct modules within the visual branch to enhance the analysis of speaker engagement in \ac{TTM} task.

\subsubsection{Head Pose Feature Extraction Module} 
Our head pose feature extraction module follows the general 6D rotation representation proposed by Zhou et al.~\cite{zhou2019continuity} and employs the RepVGG backbone~\cite{ding2021repvgg} for efficient feature extraction. Compared to these works, our contributions are twofold: (1) we integrate the 6D representation into the \textit{EgoAdapt} pipeline for the \ac{TTM} task, enabling head pose estimation to serve as an intermediate representation for multi-modal temporal reasoning; and (2) we adapt the feature extraction and normalization process to the egocentric setting, where head crops are obtained from continuous first-person video with varying scale and occlusion. In this way, our module extends~\cite{zhou2019continuity} and~\cite{ding2021repvgg} by embedding their strengths into a unified framework tailored for egocentric conversation understanding.
As shown in Fig.~\ref{fig:VSTR}, the goal of the head pose feature extraction module is to estimate the head orientation in three-dimensional space by computing the yaw (\( \alpha \)), pitch (\( \beta \)), and roll (\( \gamma \)) angles based on the input head crops \( D_h \). The first step is to normalize the image to ensure consistent scaling. This is achieved by computing \( D_{h_{norm}} = \frac{D_h - \mu}{\sigma} \), where \( \mu \) and \( \sigma \) are the mean and standard deviation of the training dataset, respectively.
Next, the normalized image \( D_{h_{norm}} \) is passed through the trained RepVGG model ~\cite{ding2021repvgg} for feature extraction, represented~as: 
\begin{equation}
 F_h = \text{RepVGG}(D_{h_{norm}}; W_{RepVGG}) 
\end{equation}
where \( F_h \in \mathbb{R}^{N_f} \) is the feature representation obtained from the model, and \( W_{RepVGG} \) are the model weights.
To efficiently predict the head pose, a 6D rotation matrix representation is employed. Following the approach by Zhou et al.~\cite{zhou2019continuity}, the rotation matrix is reduced to a 6-dimensional representation by dropping the last column vector of the rotation matrix, ensuring that the orthogonality constraint is satisfied. 
This 6D representation, denoted \( g_{GS} \), is given by:
\begin{equation}
A=g_{G S}(F_h)=\left(\left[\begin{array}{ccc}
\mid & \mid & \mid \\
a_{1} & a_{2} & a_{3} \\
\mid & \mid & \mid
\end{array}\right]\right)=\left[\begin{array}{cc}
\mid & \mid \\
a_{1} & a_{2} \\
\mid & \mid
\end{array}\right]
\end{equation}
This 6D representation is then mapped back to the 3x3 rotation matrix, denoted by \( f_{GS} \), using the following equation:
\begin{equation}
   B= f_{G S}(A)=\left(\left[\begin{array}{cc}
\mid & \mid \\
a_{1} & a_{2} \\
\mid & \mid
\end{array}\right]\right)=\left[\begin{array}{ccc}
\mid & \mid & \mid \\
b_{1} & b_{2} & b_{3} \\
\mid & \mid & \mid
\end{array}\right]
\end{equation}
The vectors \( b_1, b_2, b_3 \) are computed as follows:
\begin{equation}
\begin{aligned}
&b_{1}=\mathrm{Norm}\left(a_1\right)=\frac{a_1}{\left\|a_1\right\|}\\&b_{2}=\mathrm{Norm}\left(u_2\right)=\frac{u_2}{\left\|u_2\right\|},u_2=a_2-\left(b_1\cdot a_2\right)b_1\\&b_{3}=b_1\times b_2,
\end{aligned}
\end{equation}
where \(\text{Norm}(\cdot)\) represents a normalization function using the $L_2$-$norm$.
The orthogonal features \( B \) are used to predict the head pose angles. This is done by applying a weight matrix \( W_{out} \in \mathbb{R}^{3 \times 3N_f} \) and a bias \( b_{out} \in \mathbb{R}^{3} \), leading to the prediction of the pose angles \( \theta = \begin{bmatrix} \alpha \\ \beta \\ \gamma \end{bmatrix} \). The regression equation is given by
\begin{equation}
  \theta = W_{out} \cdot B + b_{out}   
\end{equation}
where \( \alpha \), \( \beta \), and \( \gamma \) represent the yaw, pitch, and roll angles, respectively. Afterward, the predicted angles \( \theta \) are converted into a rotation matrix \( R_p \) using the function \( R_p = \text{EulerToRotationMatrix}(\theta) \). The final output is the predicted head pose angles \( z_h = \theta \), which represent the head's orientation in three-dimensional space.

\subsubsection{Lip Feature Extraction Module} 

Let \( D_l \in \mathbb{R}^{H \times W \times C} \) denote the input image, where \( H \) and \( W \) are the height and width, and \( C \) is the number of channels (e.g., RGB). The image is partitioned into non-overlapping patches of size \( P = 32 \), resulting in \( N = \frac{H \times W}{P^2} \) patches, each flattened into a vector \( p_i' \in \mathbb{R}^{P^2 \times C} \). These flattened patches are then projected into a new feature space \( D \) via a linear projection matrix \( W_p \in \mathbb{R}^{D \times P^2 \times C} \), yielding the patch representations \( \mathbf{z}_i\).
To preserve spatial relationships between the image patches, positional encodings \( \mathbf{P}_i \in \mathbb{R}^D \) are added to each patch feature \( \mathbf{z}_i \). The final patch representation, after adding the positional encoding, is mathematically expressed as:
\begin{equation} 
\mathbf{z}_i^{\text{emb}} = \mathbf{z}_i + \mathbf{P}_i
\end{equation}
where \( \mathbf{z}_i \) is the feature vector and \( \mathbf{P}_i \) is the positional encoding for the \(i\)-th patch.
The sequence of positionally encoded patches is then represented as:
\begin{equation}
\mathbf{Z}^{\text{emb}} = [\mathbf{z}_1^{\text{emb}}, \mathbf{z}_2^{\text{emb}}, \dots, \mathbf{z}_N^{\text{emb}}]
\end{equation}
where \( \mathbf{Z}^{\text{emb}} \in \mathbb{R}^{N \times D} \) is the matrix of positionally encoded patch features.
% Each individual patch feature \( \mathbf{z}_i^{\text{emb}} \in \mathbb{R}^D \) represents the final embedded feature of the \(i\)-th patch.
To facilitate the extraction of global features, a special classification token \( \mathbf{z}_{\text{CLS}} \in \mathbb{R}^D \) is prepended to the input sequence. 
% This token is designed to capture the overall image information. 
The modified input sequence, including the [CLS] token and the positionally encoded patch features, is represented as:
\begin{equation}
\mathbf{Z}^{\text{input}} = [\mathbf{z}_{\text{CLS}}, \mathbf{z}_1^{\text{emb}}, \dots, \mathbf{z}_N^{\text{emb}}]
\end{equation}
where \( \mathbf{z}_{\text{CLS}} \) is the global feature token and each \( \mathbf{z}_i^{\text{emb}} \) is the final encoded representation of the \(i\)-th patch after the addition of positional encodings.
% This sequence is then input into the Transformer encoder. The Transformer encoder processes the sequence through its self-attention mechanism, capturing dependencies between patches. 
The output of the Transformer encoder is the matrix \( \mathbf{Z}^{\text{enc}} \), defined as:
\begin{equation}
\mathbf{Z}^{\text{enc}} = \text{Transformer Encoder}(\mathbf{Z}^{\text{input}})
\end{equation}
where \( \mathbf{Z}^{\text{enc}} \in \mathbb{R}^{(N+1) \times D} \) represents the processed sequence of patch features.
% Each element \( \mathbf{z}_i^{\text{enc}} \in \mathbb{R}^D \) corresponds to the final representation of the \(i\)-th patch after self-attention.
The output corresponding to the [CLS] token, \( \mathbf{z}_{\text{CLS}}^{\text{out}} \), serves as the global feature vector \( z_l \) for the entire image. This global feature vector is defined as:
\begin{equation}
    z_l = \mathbf{z}_{\text{CLS}}^{\text{out}} \in \mathbb{R}^D
\end{equation}
where \( z_l \in \mathbb{R}^D \) represents the high-dimensional semantic representation of the image.
% , encapsulating its global features after the self-attention mechanism has processed the dependencies between patches.

\subsection{Parallel Shared-weight Audio Encoder}
\begin{figure*}[!tb]
\centering
\includegraphics[width=0.85\textwidth]{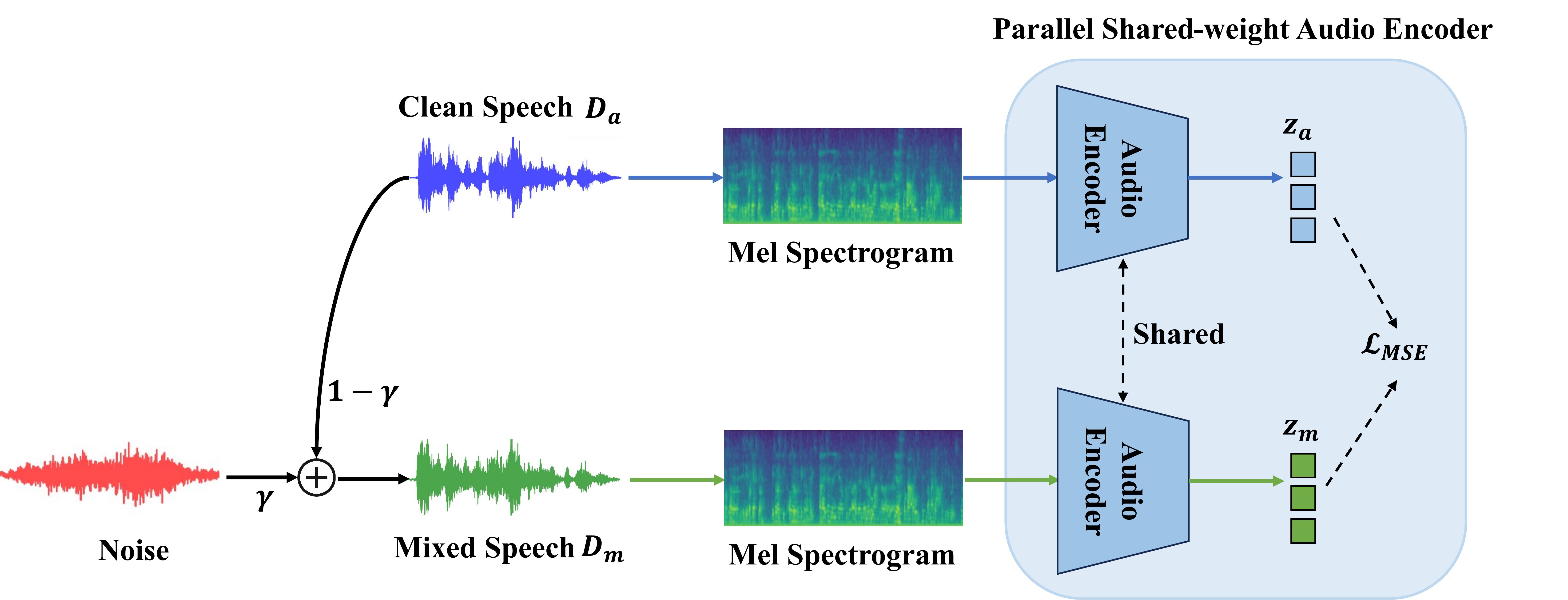} 
\caption{\acf{PSA} encoder. It processes clean speech $D_a$ and noise-mixed speech $D_m$ (ratio $\gamma$) into Mel spectrograms, encodes them with shared weights to obtain embeddings $z_a$ and $z_m$, and uses MSE loss to enforce noise-robust audio features.}
\label{fig:PSA}
\end{figure*}

To enhance noise robustness, we propose a special audio encoder as shown in Fig.~\ref{fig:PSA}.
Let \( D_a(t) \) denotes the clean speech signal, \( n(t) \) represents the additive noise, and \( D_m(t) \) be the mixed speech, expressed as:
\begin{equation}
   D_m(t) = (1 - \gamma) \cdot D_a(t) +  \gamma \cdot n(t)
\end{equation}
where \( \gamma \) is a weighting factor that controls the proportion of noise to clean speech.
Then, clean speech $D_a$ and mixed speech $D_m$ are transformed into their respective Mel spectrograms: \( S_a = \text{MelSpec}(D_a) \) and  \(S_m = \text{MelSpec}(D_m) \), where \( \text{MelSpec}(\cdot) \) denotes  the Mel spectrogram function that provides a time-frequency representation suitable for audio signal processing. The audio encoder \( f_{enc} \) operates on both the clean and mixed Mel spectrograms to derive latent embeddings \( z_a = f_{enc}(S_a; \theta_a) \quad \text{and} \quad z_m = f_{enc}(S_m; \theta_m) \) where \( \theta_a \) and \( \theta_m \) are the parameters of the encoder associated with clean and mixed speech, respectively. 
The shared loss function, \( \mathcal{L}_{MSE} \), is defined as the squared Euclidean distance between the latent embeddings of clean and mixed speech ( \( z_a \) and \( z_m \) ), given by:
\begin{equation}
    \mathcal{L}_{MSE} = || z_a - z_m ||^2
\end{equation}
This encourages the encoder to extract noise-robust features by training the model to learn representations from both clean and noisy inputs~\cite{ephraim2006recent, benesty2006speech}, pushing the model to focus on the essential characteristics of the speech signal while disregarding noise components.

For audio encoder, we adopt the Whisper-small model~\cite{radford2021learning}, which has been pre-trained on over 680,000 hours of diverse speech data for \acf{ASR} and speech translation tasks to encode the acoustic and semantic characteristics inherent in the speech input. 
Compared to other unsupervised speech pre-trained models like Wav2vec 2.0~\cite{baevski2020wav2vec} and HuBERT~\cite{hsu2021hubert}, Whisper’s output features are more closely aligned with different downstream tasks, making it an ideal choice for task transfer learning. Furthermore, the extensive training set of Whisper enhances its performance and generalizability. 
Therefore, we select it as the Audio encoder to generate a comprehensive feature vector, denoted as \( z_a \).

\subsection{Visual Modality Missing Awareness Module}
\begin{figure*}[!tb]
\centering
\includegraphics[width=0.75\textwidth]{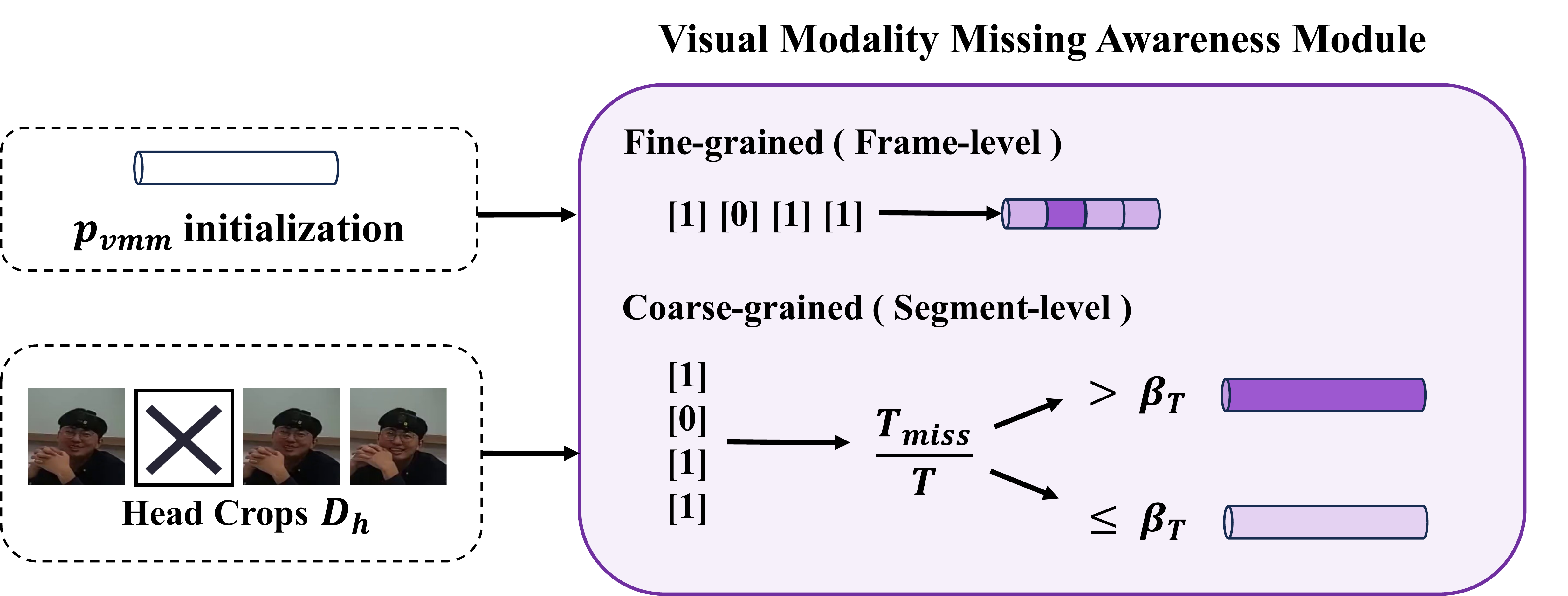} 
\caption{\acf{VMMA} module dynamically handles missing visual inputs by processing head crops~\( D_h \) and the initialized prompt \( p_{\text{vmm}} \). It extracts fine-grained features by tracking frame presence and coarse-grained features by quantifying the missing frame ratio \( \Delta_c \).
 }
\label{fig:VMMA}
\end{figure*}

The \acf{VMMA} module (Fig.~\ref{fig:VMMA}),  is designed to address the challenges associated with missing visual information i.e., Fig.~\ref{fig:task and challenges}~(c). 
The module processes two different inputs: head crops \(D_h\), and \( p_{\text{vmm}}\) initialization. 
The inputs undergo a transformation into a Visual Mask Matrix that is dynamically updated to reflect the presence or absence of visual cues, which extracts both \textit{coarse-grained} and \textit{fine-grained} features.

\subsubsection{Fine-grained}
Firstly, our \ac{VMMA} prompt \( p_{\text{vmm}}^{b, t, d} \) is initialized to zero to establish a foundation for integrating modality-specific details into the model. 
In this context, \( b, t, d \) represents the batch size, frame number, and feature dimension. This initialization allows for dynamic adjustments during training, effectively managing instances where modalities are missing.

The fine-grained prompt \( p_{\text{vmm}}^{b, t, d} \) focuses explicitly on the temporal dimension \( T \) by indicating whether specific frames are present or missing. 
If a head crop at \( t \) is present, \( p_{\text{vmm}}^{b, t, d} \) is set to zero; otherwise \( p_{\text{vmm}}^{b, t, d} \) takes a value of one during visual absence.
This mechanism efficiently adjusts the values of tensor \( p_{\text{vmm}}^{b, t, d} \) based on frame availability, significantly enhancing the model's ability to manage and rectify missing modalities with precision.

\subsubsection{Coarse-grained}
To quantify the absence of frames within a head crop, we introduce \( \Delta_c \), defined as the ratio of missing frames to the total frame number: 

\begin{equation}
\Delta_c = \frac{T_\text{miss}}{T}
\end{equation}
where \( T_\text{miss} \) represents the number of missing head crops and \( T \) is the total number of frames. By setting a threshold \( \beta_T \) to gauge the severity of missing data, \( p_{\text{vmm}}^{b, t, d} \) is modified accordingly

\begin{equation}
p_{\text{vmm}}^{b, t, d} = 
\begin{cases} 
1 & \text{if }(\Delta_c < \beta_T \text{ and } (d \leq \frac{D}{2}) ) \text{ or } ( \Delta_c \geq \beta_T  \text{ and }(d > \frac{D}{2}))\\ 
0 & \text{otherwise}
\end{cases}
\end{equation}

This formula delineates how \( p_{\text{vmm}}^{b, t, d} \) is assigned a value of 1 or 0 in different segments of the feature dimension \( D \), depending on whether the proportion of missing frames \( \Delta_c \) falls below or exceeds the threshold \( \beta_T \). 
Specifically, when the proportion of missing data is less than \( \beta_T \), the first half of the feature dimensions is activated (set to 1), while the second half is deactivated (set to 0). 
In contrast, when the proportion of missing data surpasses \( \beta_T \), the first half is deactivated and the second half is activated. 
This adaptive mechanism enables the model to effectively respond to varying levels of missing data in video frames.

To enhance the \ac{VMMA} module effectiveness, we propose a dynamic design for the threshold \( \beta_T \), which is adaptable based on the data context and the operational environment. 
It can be fine-tuned using a validation dataset, allowing the model to learn from previous instances of missing data. The adaptive threshold is formulated as
\begin{equation}
\beta_T = \text{mean}(\Delta_c) + k \cdot \text{std}(\Delta_c)
\end{equation}
where \( k \) is a tunable parameter that adjusts the sensitivity of the threshold based on the distribution of missing data across different scenarios. 

\subsection{Visual-Audio Fusion Module}
\begin{figure*}[!tb]
\centering
\includegraphics[width=0.6\textwidth]{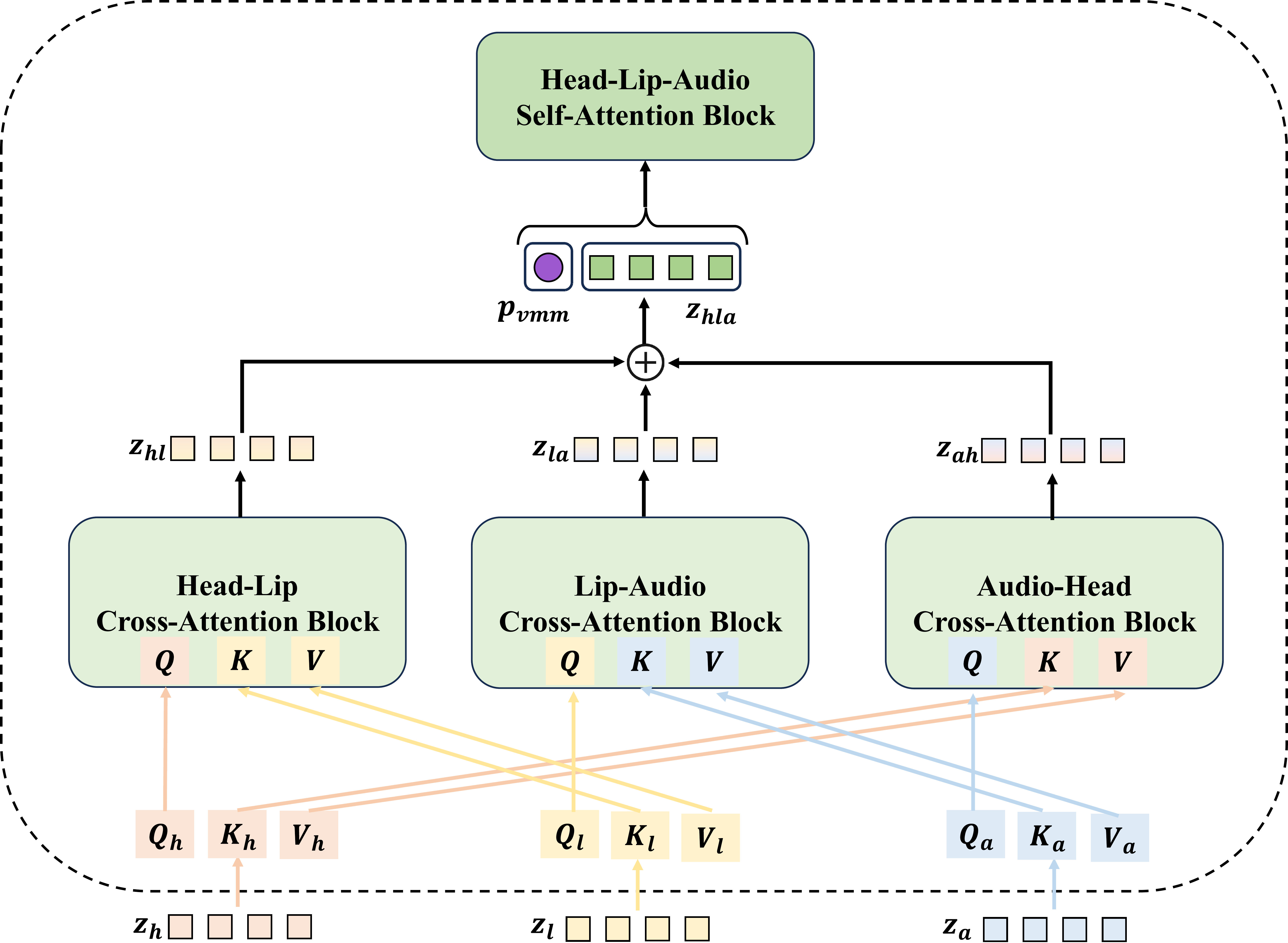} 
\caption{Oveview of Visual-Audio Fusion module. It fuses head pose $z_h$, lip motion $z_l$, and audio $z_a$ via Head-Lip, Lip-Audio, and Audio-Head cross-attention blocks. Aggregated features with VMMA prompt $p_{\mathrm{vmm}}$ are refined by self-attention for final ``Talking to Me'' prediction.
}
\label{fig:Fusion}
\end{figure*}

In the \ac{TTM} task, integrating information from different modalities can be carried out in various stages. Among existing methods, the Ego4D baseline for \ac{TTM}~\cite{grauman2022ego4d} and EgoT2 ~\cite{xue2023egocentric} employed feature-level fusion in which the extracted features of each modality (e.g., visual and audio) are fused early, before being input to a shared classification model. 
Although this fusion method allows the model to learn correlations between the modalities at the feature level, it can struggle with misaligned or noisy data. 
QuAVF ~\cite{lin2023quavf} applied decision-level fusion in which each modality has its own classification model, and the results of each model are combined to make the final decision. This fusion method is more robust to errors in individual modalities but may miss fine-grained interactions.
We considered multiple possible fusion configurations among the three modalities. 
The chosen design (Head–Lip, Lip–Audio, Audio–Head) was motivated by (i) physiological and conversational relevance (lip motion tightly couples with audio; head pose naturally correlates with both speech rhythm and lip dynamics), and (ii) empirical performance. Alternative pairings such as Lip–Head or Audio–Lip–Head chains were tested in preliminary experiments but showed marginal or redundant gains compared with the selected combination. Hence, we retained the current setup as a balanced trade-off between model complexity and effectiveness. 

As illustrated in Fig.~\ref{fig:Fusion}, we propose a Visual-Audio Fusion module that integrates multi-modal inputs derived from three primary sources: 1) the head pose feature extraction module, which generates the head features represented as \( z_h \); 2) the lip feature extraction module, which produces the lip features denoted as \( z_l \) and 3) the \ac{PSA} encoder, which extracts audio features represented as \( z_a \). Following the generation of the input features, the module employs specially designed two-by-pair interactive cross-attention mechanisms to enhance the interactions between modalities. We illustrate this using the head-lip cross-attention block as an example.
In this block, the head features \( z_h \) are used as queries \( Q \), while the lip features \( z_l \) serve as keys \( K \) and values \( V \). The attention mechanism is computed as follows
\begin{equation}
 z_{hl} = \text{Softmax}\left(\frac{Q_h K_l^T}{\sqrt{d_k}}\right) V_l
\end{equation}
where \( d_k \) represents the key vectors' dimensionality, the softmax function normalizes the attention scores. The resulting output \( z_{hl} \) encapsulates the enhanced representation of head and lip interactions. Similarly, the lip-audio cross-attention block and the audio-head cross-attention block are constructed using the same principles. For the Lip-Audio Cross-Attention block, lip features \( z_l \) are used as queries, while audio features \( z_a \) are used as keys and values, resulting in output \( z_{la} \). For the Audio-Head Cross-Attention block, the audio features \( z_a \) serve as queries, the head features \( z_h \) acting as keys and values, resulting in the output \( z_{ah} \). The outputs from the three cross-attention blocks, \( z_{hl} \), \( z_{la} \), and \( z_{ah} \), are then aggregated through an addition operation to form a unified representation
\begin{equation}
z_{hla} = z_{hl} + z_{la} + z_{ah}
\end{equation}
The aggregated representation \( z_{hla} \) is concatenated with the output \( p_{\text{vmm}} \) obtained from the \ac{VMMA} module. This concatenated representation is then fed into the Head-Lip-Audio Self-Attention block, which further processes the combined information. This output culminates in the \ac{TTM} prediction results, enabling a comprehensive understanding of the multi-modal interactions.

\section{EXPERIMENTATION, RESULT AND ANALYSIS}
\subsection{Datasets}
We conducted our experiments using the Ego4D social interaction benchmark~\cite{grauman2022ego4d}, a pivotal dataset in social interaction research that captures complex, real-world interactions from an egocentric perspective. This dataset comprises 389 training clips, amounting to approximately 32.4 hours of footage, making it suitable for training models to recognize and interpret social cues in dynamic environments. It includes 50 validation clips with around 4.2 hours and 133 test clips totaling 11.1 hours. The video data are recorded at a frame rate of 30 frames per second, while the audio is sampled at 16 kHz, ensuring high-quality input for our analysis.

To enhance the noise robustness of our system, we selected the WHAM! dataset~\cite{wichern2019wham} as the source of noise data. This dataset is crucial for assessing the model's performance in real-world scenarios, as it consists of noisy mixtures generated by combining speech from the WSJ0-2mix corpus with real-world noise recordings, providing approximately 42 hours of noisy speech data. This diversity in noise conditions is essential for training robust models capable of performing well in varied auditory environments. To adapt this dataset to our task, we randomly select a noise sequence from the WHAM! dataset with a 50\% probability and add it to the original clean speech, thereby creating mixed speech for training purposes. We also experimented with various \acf{SNR} ranging from -10 to 10 dB to evaluate performance under different noise conditions.

\subsection{Baselines}
We compare with four representative methods:  

(1) Random Guess~\cite{grauman2022ego4d}: outputs two-label predictions via Bernoulli sampling.  

(2) ResNet-18 Bi-LSTM~\cite{grauman2022ego4d}: extracts facial and audio embeddings with ResNet-18 and ResNet-SE, concatenates them, and predicts \ac{TTM} with a fully-connected layer.  

(3) EgoT2~\cite{xue2023egocentric}: refines outputs from models trained for separate tasks.  

(4) QuAVF~\cite{lin2023quavf}: filters data via face quality scores, encodes visual/audio features with ResNet-50~\cite{he2016deep} and Whisper~\cite{radford2021learning}, and applies quality-aware fusion.  

All methods are evaluated under identical parameter settings.

\subsection{Evaluation Metrics}
Following previous works~\cite{grauman2022ego4d, xue2023egocentric, lin2023quavf}, we use the \ac{mAP}, which measures the precision-recall trade-off, and Top-1 accuracy, which indicates the proportion of correctly predicted top-ranking results, as evaluation metrics (the higher value corresponds to better results). 
Unless otherwise specified, we conduct comparative experiments in the test set and the validation set, while ablation experiments are performed on the validation set.

\subsection{Backbones}

\begin{table*}[!tb]
\centering
\caption{Head pose feature extraction parameters.}
\begin{tabular}{l|l}
\toprule
\textbf{Parameter} & \textbf{Head Pose Feature Extraction} \\
\midrule
Backbone Model & 6DRepNet \\
Pretrained & Yes (frozen weights)\\
Deploy Mode & True \\
Input Feature & 224x224 pixels image \\
Pooling Layer & Adaptive Average Pooling \\
Output Dimension & 6 (Ortho6D Representation) \\
Rotation Matrix Computation & Ortho6D to Rotation Matrix \\
Euler Angles Computation & Rotation Matrix to Euler Angles \\
Training Dataset & 300W-LP \\
\bottomrule
\end{tabular}
\label{tab:head_params}
\end{table*}

\subsubsection{Head Pose Feature Extraction}
According to Table~\ref{tab:head_params}, we use 6DRepNet~\cite{hempel20226d} with frozen pretrained weights for head pose estimation. 
The model runs in deploy mode on 224x224 images, applies adaptive average pooling, and outputs a 6D Ortho6D representation~\cite{andreou2021hierarchy,andreou2022pose}. 
Rotation matrices are computed from Ortho6D and converted to Euler angles. 
The backbone is trained on the 300W-LP dataset~\cite{zhu2016face}.

\begin{table*}[!tb]
\centering
\caption{Lip and audio feature extraction parameters.}
\begin{tabular}{l|l|l}
\hline
\textbf{Parameter} &\textbf{ Lip Feature Extraction} & \textbf{Audio Feature Extraction} \\
\hline
Backbone Model & CLIP ViT-B/32 & Whisper-small \\
Pretrained & Yes (frozen weights) & Yes (frozen weights) \\
Input Feature & 224x224 pixels image & Log-Mel Spectrogram \\
Embedding Dimension & 512 & 768 \\
Temporal Module Dimension & 128 & 128 \\
Number of Temporal Layers & 1 & 1 \\
Number of Attention Heads & 8 & 8 \\
Max Temporal Length & 900 & 1500 \\
Position Embedding & Learnable, max length: 900 & Learnable, max length: 1500 \\
Classification Token (CLS) & Yes & N/A \\
Dropout & 0 & 0.25 \\
\hline
\end{tabular}
\label{tab:ve_ae_params}
\end{table*}

\subsubsection{Lip Feature Extraction}
As shown in Table~\ref{tab:ve_ae_params}, we use CLIP Vision Transformer (ViT)-B/32~\cite{radford2021learning} to encode 224x224 lip images into 512-d features. 
A temporal module (dim 128, 1 layer, 8 heads) processes sequences of up to 900 frames with learnable position embeddings and a CLS token. 
No dropout is applied.

\subsubsection{Audio Feature Extraction}
As shown in Table~\ref{tab:ve_ae_params}, we use Whisper-small~\cite{radford2021learning} to encode log-Mel spectrograms into 768-d features. 
The temporal module matches the lip branch but handles sequences up to 1500 frames. 
\subsection{Training Settings}
We train with Adam (lr $1\times10^{-5}$) and Focal Loss ($\gamma=2$, $\alpha=0.25$). 
Training runs for 40 epochs with gradient clipping (1.0) and early stopping (patience 100). 
Batch size is 1 with 4 data loader workers. 
Head and lip images are resized to $224\times224$; audio is represented by 80-d log-Mel features (25 ms window, 10 ms hop, 16 kHz). 
Noise augmentation (ratio 0.5) uses WHAM! noise with 50\% probability. 
A coarse-grained prompt type (threshold 0.2) is adopted. 
Cross-attention has 6 randomly initialized multi-head layers (8 heads each); self-attention~\cite{tao2021someone} has 1 layer with 8 heads.

\begin{table*}[!tb]  
\centering  
\caption{Comparison of our proposed EgoAdapt with existing methods  on \ac{TTM} benchmark of Ego4D dataset. }  
\begin{tabular}{lcccc}  
\hline  
\textbf{Method} & \multicolumn{2}{c}{\textbf{Val}} & \multicolumn{2}{c}{\textbf{Test}}  \\
       & \ac{Acc}(\%) & \ac{mAP}(\%)& \ac{Acc}(\%) & \ac{mAP}(\%) \\
\hline  
Random Guess ~\cite{grauman2022ego4d}&33.44&53.82& 47.41& 50.16\\
ResNet-18 Bi-LSTM ~\cite{grauman2022ego4d}&64.31&56.50& 49.75& 55.06\\
EgoT2 ~\cite{xue2023egocentric}&66.07&60.80& 56.17& 57.82\\
QuAVF ~\cite{lin2023quavf}&71.80&71.20& 57.05& 65.83\\
Ours&\textbf{81.14}&\textbf{84.83}&\textbf{62.01}& \textbf{67.39}\\
\hline  
\end{tabular}  
\label{tab:sota}  
\end{table*}

% \begin{figure}[!tb]
%     \centering
%     \begin{minipage}[t]{0.32\textwidth}
%         \centering
%         \includegraphics[width=\textwidth]{img/PCA-resnet18.pdf}
%         \caption*{(a) ResNet-18 Bi-LSTM ~\cite{grauman2022ego4d}}
%     \end{minipage}
%     \hfill
%     \begin{minipage}[t]{0.32\textwidth}
%         \centering
%         \includegraphics[width=\textwidth]{img/PCA-QuAVF.pdf}
%         \caption*{(b) QuAVF ~\cite{lin2023quavf}}
%     \end{minipage}
%     \hfill
%     \begin{minipage}[t]{0.32\textwidth}
%         \centering
%         \includegraphics[width=\textwidth]{img/PCA-ours.pdf}
%         \caption*{(c) Ours}
%     \end{minipage}
%     \caption{Feature space visualization using (a) ResNet-18 Bi-LSTM, (b) QuAVF and (c) our method. The green points represent instances of "Talking to Me" (positive class), and the red points represent "Non Talking to Me" (negative class). 
%     }
%     \label{fig:pca}
% \end{figure}

\subsection{Comparison with the State-of-the-art}
The experimental results are presented in Table~\ref{tab:sota}.
Previous approaches such as Random Guess ~\cite{grauman2022ego4d} and ResNet-18 Bi-LSTM ~\cite{grauman2022ego4d} focus mainly on basic classification tasks and simple fusion of multimodal data. 
These methods either ignore or inadequately leverage the intricate cooperation between audio and visual modalities, which is crucial for a more comprehensive understanding of egocentric video content. 
For example, Random Guess achieves only 47.41\% \ac{Acc} and 50.16\% \ac{mAP} on the test set, while ResNet-18 Bi-LSTM slightly improves the results to 49.75\% \ac{Acc} and 55.06\% \ac{mAP}, yet still falls short in handling the complex interplay between modalities. 
While some methods, such as EgoT2 ~\cite{xue2023egocentric} and QuAVF ~\cite{lin2023quavf}, attempt to incorporate both audio and visual signals, they have poor performance compared to EgoAdapt. 
Specifically, EgoT2 achieves 55.93\% \ac{Acc} and 57.51\% \ac{mAP} on the test set, while QuAVF reaches 57.05\% \ac{Acc} and 65.83\% \ac{mAP}. 
However, these methods are primarily designed for tasks involving trimmed short videos, limiting their ability to model long-range dependencies and interactions necessary for egocentric interactive speaker detection. 
QuAVF, despite its reasonable results, has constrained performance due to its reliance on static quality filtering and less dynamic fusion strategies.
Our proposed EgoAdapt outperforms all the comparison methods with the highest scores in both the validation set (77.40\% \ac{Acc}, 79.37\% \ac{mAP}) and the test set (61.95\% \ac{Acc}, 66.75\% \ac{mAP}). 
As a result, EgoAdapt sets a new benchmark in egocentric interactive speaker detection, outperforming all previous methods by a significant margin.

\subsection{Ablation Study}
\begin{table*}[!tb]  
\centering  
\caption{
 Comparison of the performance for different modality and backbone combinations.
 }
\begin{tabular}{ccccl}
\hline
\textbf{Modality}  & \textbf{Visual Backbone} & \textbf{Audio Backbone} &\textbf{\ac{Acc}(\%) } & \textbf{\ac{mAP}(\%)} \\
\hline
V& ViT ~\cite{radford2021learning} & \ding{55}  &67.25 & 62.51 \\
A & \ding{55}&Whisper ~\cite{radford2021learning}&\textbf{75.64} & \textbf{77.55} \\
V, A& ViT & WavLM ~\cite{chen2022wavlm}&72.69 & 72.96 \\
V, A& TalkNet ~\cite{tao2021someone}& WavLM&72.52 & 72.59 \\
V, A& TalkNet& Whisper&73.65 & 77.51\\
V, A& ViT & Whisper&75.33 & 77.21 \\

\hline
\end{tabular}
\label{tab:backbone}
\end{table*}
\subsubsection{Preliminary study on modality/backbone combinations }
Before presenting our proposed Visual-Audio Fusion module, we first conduct preliminary experiments on different modality and backbone combinations, following the experimental setup of QuAVF~\cite{lin2023quavf} for unimodal and TalkNet~\cite{tao2021someone} for multimodal. The results are summarized in Table~\ref{tab:backbone}.  
% \subsubsection{Impact Assessment of different modality and backbone combinations}
% Before presenting our proposed Visual-Audio Fusion module, we evaluate different modality and backbone combinations as preliminary studies (Table 4)
% Table~\ref{tab:backbone}  presents the results of our preliminary experiments on various combinations of modalities and backbone networks, conducted in accordance with the experimental setup described in QuAVF~\cite{lin2023quavf} for unimodal and TalkNet~\cite{tao2021someone} for multimodal. 
Notably, our results align with those reported in ~\cite{lin2023quavf}, as the unimodal A consistently outperforms the multimodal A+V, even when different backbone networks are employed.
For example, when using the Whisper audio backbone for modality A, the unimodal performance achieves an \ac{Acc} of 75.64\% and an \ac{mAP} of 77.55\%. 
In comparison, the best performing multimodal combination (A+V with ViT visual backbone and Whisper audio backbone) results in an \ac{Acc} of 75.33\% and an \ac{mAP} of 77.21\%. The unimodal A outperforms the multimodal A+V by 0.31\% in \ac{Acc} and 0.34\% in \ac{mAP}.
Additionally, the performance of the unimodal visual modality (V) is significantly lower than the unimodal audio modality (A). 
Specifically, the \ac{Acc} of V with the ViT backbone is 67.25\%, while the unimodal audio modality Whisper achieves 75.64\%, an 8.39\% increase in \ac{Acc}. 
Even in multimodal combinations (V, A), such as ViT + Whisper, the performance is still inferior to the unimodal audio results, highlighting the limited contribution of the visual modality in this setup. 

The above analysis suggests that, in this particular experimental setup, the addition of the visual modality does not provide a substantial advantage and, in some cases, may even slightly hinder performance. 
We identify two primary factors contributing to this phenomenon:   
(1) Data-related issues: Approximately one-third of the head crops in the dataset are missing, leading to incomplete visual data that impedes the model's ability to effectively leverage visual features.
(2) Model-related issues: Previous models primarily focused on identifying speaking states but did not delve into the specifics of the speaking target, resulting in misclassifications, particularly errors in labeling speakers who are not speaking to the camera-wearer. 
Our research seeks to address these challenges and provide a more effective A+V solution for the TTM task, with the potential to surpass the performance of the unimodal A approach. We use the A+V configuration with the ViT visual backbone and Whisper audio backbone as our baseline.

\begin{table}[!tb]
\centering
\begin{minipage}{0.55\linewidth}
\centering
\caption{The effects of different \ac{VMMA} module settings on the validation set. $\beta$ represents an adaptive parameter trained to optimize the model's performance under the Coarse-Grained setting. (Bold: best results)}
\begin{tabular}{c|c|cc}
\hline
\textbf{\( p_{\text{vmm}}\) } & \textbf{Threshold} ($\beta_T$) & \textbf{\ac{Acc} (\%)} & \textbf{\ac{mAP} (\%)} \\
\hline
None & \ding{55} & 77.40 & 79.37 \\
\hline
Fine-Grained & \ding{55} & \textbf{81.66} & 84.53 \\  
\hline
\multirow{4}{*}{Coarse-Grained} 
& $\beta$ & 81.14 & \textbf{84.83} \\  
& 20\% & 80.29 & 84.07 \\  
& 50\% & 78.18 & 83.73 \\  
& 70\% & 78.01 & 83.50 \\  
\hline
\end{tabular}
\label{tab:abl-VMMA}
\end{minipage}
\hfill
\begin{minipage}{0.4\linewidth}
\centering
\caption{The effects of \ac{VSTR}, \ac{PSA}, and \ac{VMMA}. (Bold: best results)}
\begin{tabular}{c|c|c|c|c}
\hline
\textbf{\ac{VSTR}} & \textbf{\ac{PSA}} & \textbf{\ac{VMMA}} & \textbf{\ac{Acc}(\%)} & \textbf{\ac{mAP}(\%)} \\
\hline
\ding{55} & \ding{55} & \ding{55} & 75.33 & 77.21 \\
\ding{52} & \ding{55} & \ding{55} & 78.31 & 82.60 \\  
\ding{55} & \ding{52} & \ding{55} & 77.26 & 80.15 \\  
\ding{55} & \ding{55} & \ding{52} & 78.36 & 82.44 \\  
\ding{55} & \ding{52} & \ding{52} & 79.55 & 83.06 \\  
\ding{52} & \ding{55} & \ding{52} & 80.67 & 83.82 \\  
\ding{52} & \ding{52} & \ding{55} & 78.57 & 83.99 \\  
\ding{52} & \ding{52} & \ding{52} & \textbf{81.14} & \textbf{84.83} \\  
\hline
\end{tabular}
\label{tab:abl-all}
\end{minipage}
\end{table}

\subsubsection{Effects of the \ac{VSTR}, \ac{PSA}, and \ac{VMMA}}
Table~\ref{tab:abl-all} presents the ablation results for the \ac{VSTR} module, \ac{PSA} encoder, and \ac{VMMA} module. 
The full configuration with all three components achieves the best performance, reaching 81.14\% accuracy and 84.83\% mAP.  
When using partial configurations, \ac{VSTR}+\ac{VMMA} attains 80.67\% / 83.82\% (Acc / mAP), slightly outperforming \ac{PSA}+\ac{VMMA} (79.55\% / 83.06\%), indicating the importance of spatial-temporal reasoning in the fusion process.  
The lowest result (75.33\% / 77.21\%) is obtained when all three modules are removed. In this setting, the system relies solely on optimized backbone networks and the cross-modal fusion mechanism.  
Even under this minimal configuration, the performance already surpasses QuAVF (Table~1) by a notable margin, underscoring the strong contribution of the improved backbones (Table~4) and the effectiveness of our fusion strategy. The proposed modules then provide additional complementary gains on top of this strong baseline.

\begin{figure}[!tb]
    \centering
    \includegraphics[width=0.98\textwidth]{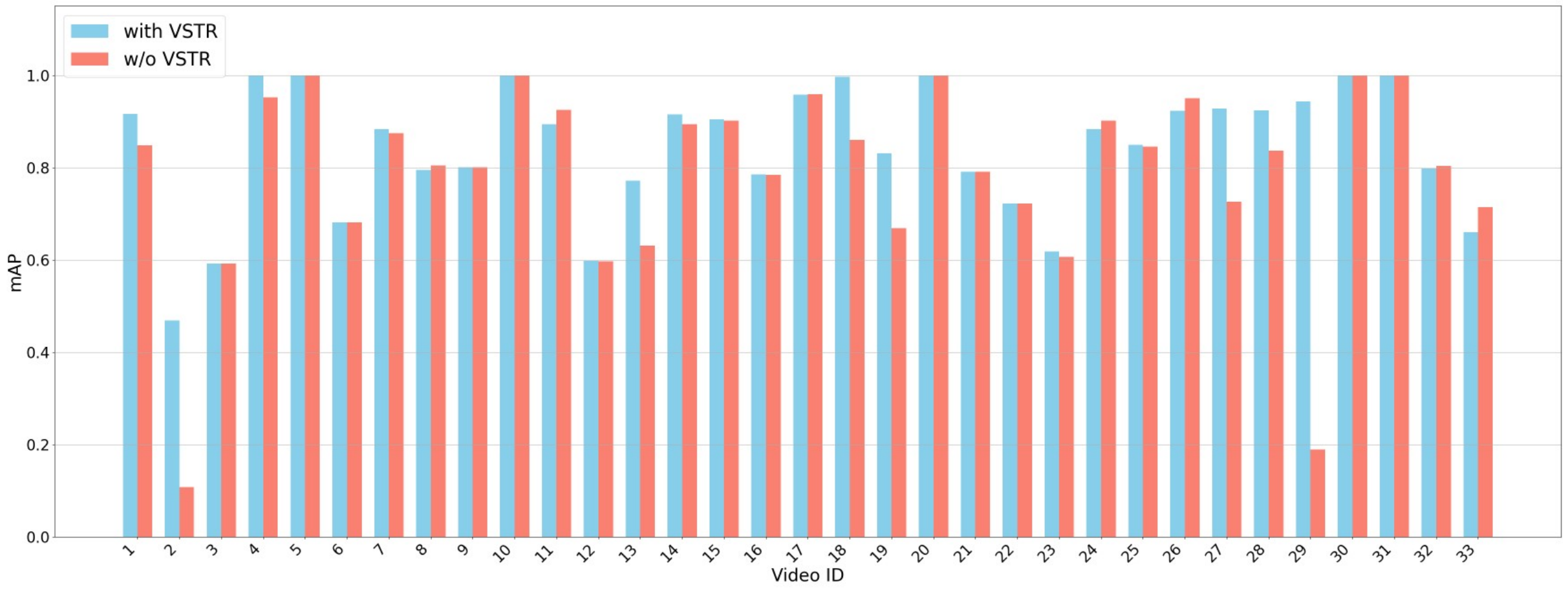}
    \caption{Comparison of the performance with or w/o VSTR.}
    \label{fig:abl-VSTR}
\end{figure}

\subsubsection{\ac{VSTR} Impact Assessment}

% As shown in Fig~\ref{fig:dataset}, we first analyzed the distribution of the dataset and found that the proportion of frames with multiple speakers in the validation set is substantial. 
The training dataset consists of approximately 32.9\% single-person frames, 31.4\% two-person frames, and 35.7\% frames with more than two persons, indicating a relatively balanced distribution across different interaction complexities. 
In contrast, the validation dataset is highly imbalanced, with 93.7\% of frames containing exactly two persons, while only 4.5\% are single-person frames and 1.8\% involve more than two persons. 
Consequently, we extracted a series of representative videos from the validation set to evaluate the effectiveness of the \ac{VSTR} module. 
Fig ~\ref{fig:abl-VSTR} compares the \ac{mAP} for different video IDs, assessing the model's performance with and without the integration of \ac{VSTR} module.
The blue bars represent the performance of our proposed method, which includes the \ac{VSTR} module, while the red bars indicate the performance when the \ac{VSTR} module is replaced by the corresponding baseline configuration.
The comparison clearly indicates that, in most video instances, the incorporation of the \ac{VSTR} module results in a marked improvement in \ac{mAP}. For example, in video IDs 1, 3, and 8, the model with \ac{VSTR} (blue bars) demonstrates a substantial increase in \ac{mAP} compared to the model without \ac{VSTR} (red bars). On average, the inclusion of the \ac{VSTR} module leads to an improvement in \ac{mAP} by approximately 0.1 to 0.15, underscoring the positive impact of this module on model performance.
These findings suggest that the \ac{VSTR} module plays a critical role in enhancing the model's ability to accurately detect and classify speakers, particularly in interactive settings with multiple speakers. 

\begin{figure*}[!tb]
    \centering
    \begin{minipage}[t]{0.47\textwidth}
        \centering
        \includegraphics[width=\textwidth]{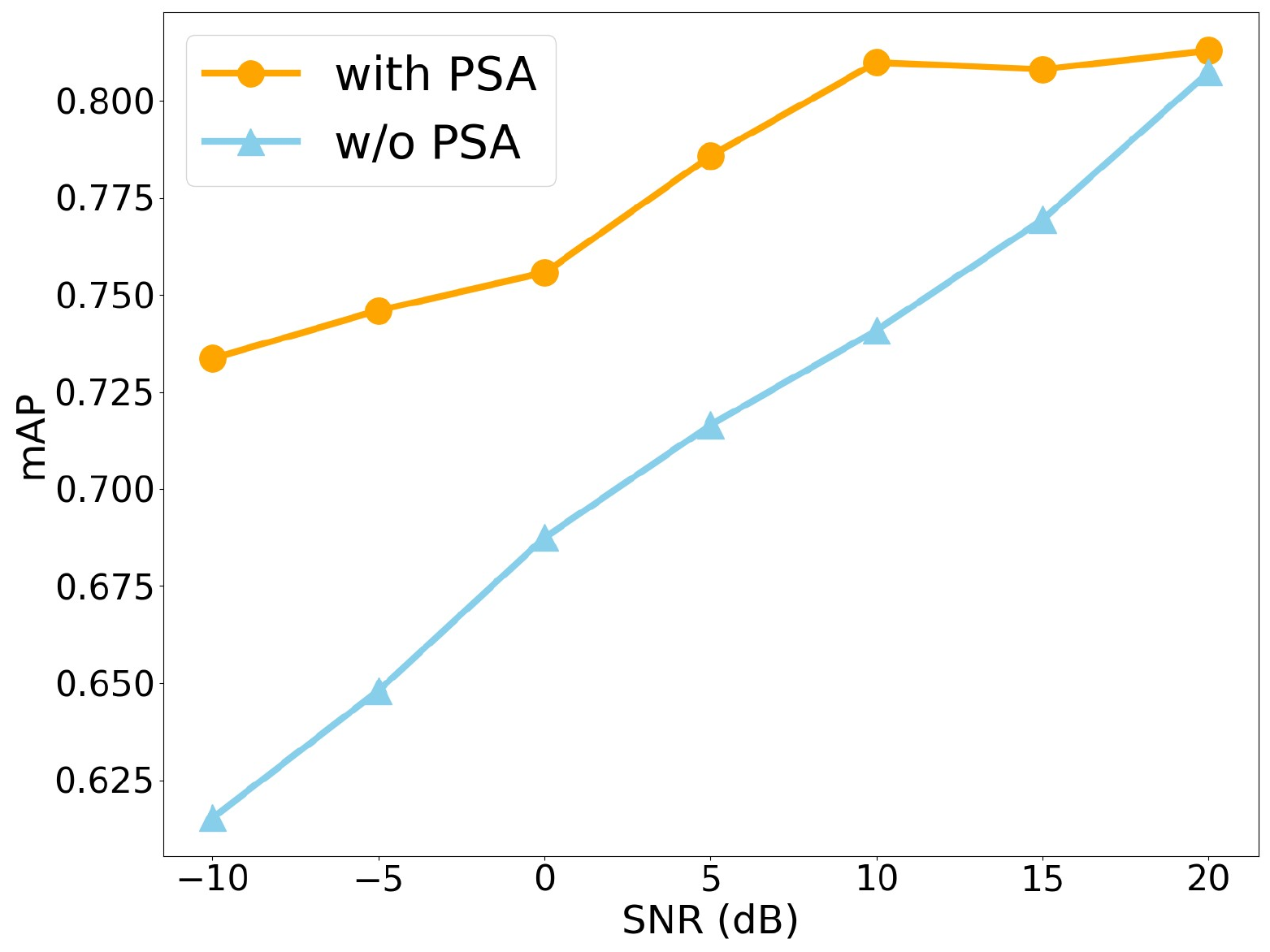}
        \caption*{(a) \ac{mAP} }
    \end{minipage}
    \hfill
    \begin{minipage}[t]{0.47\textwidth}
        \centering
        \includegraphics[width=\textwidth]{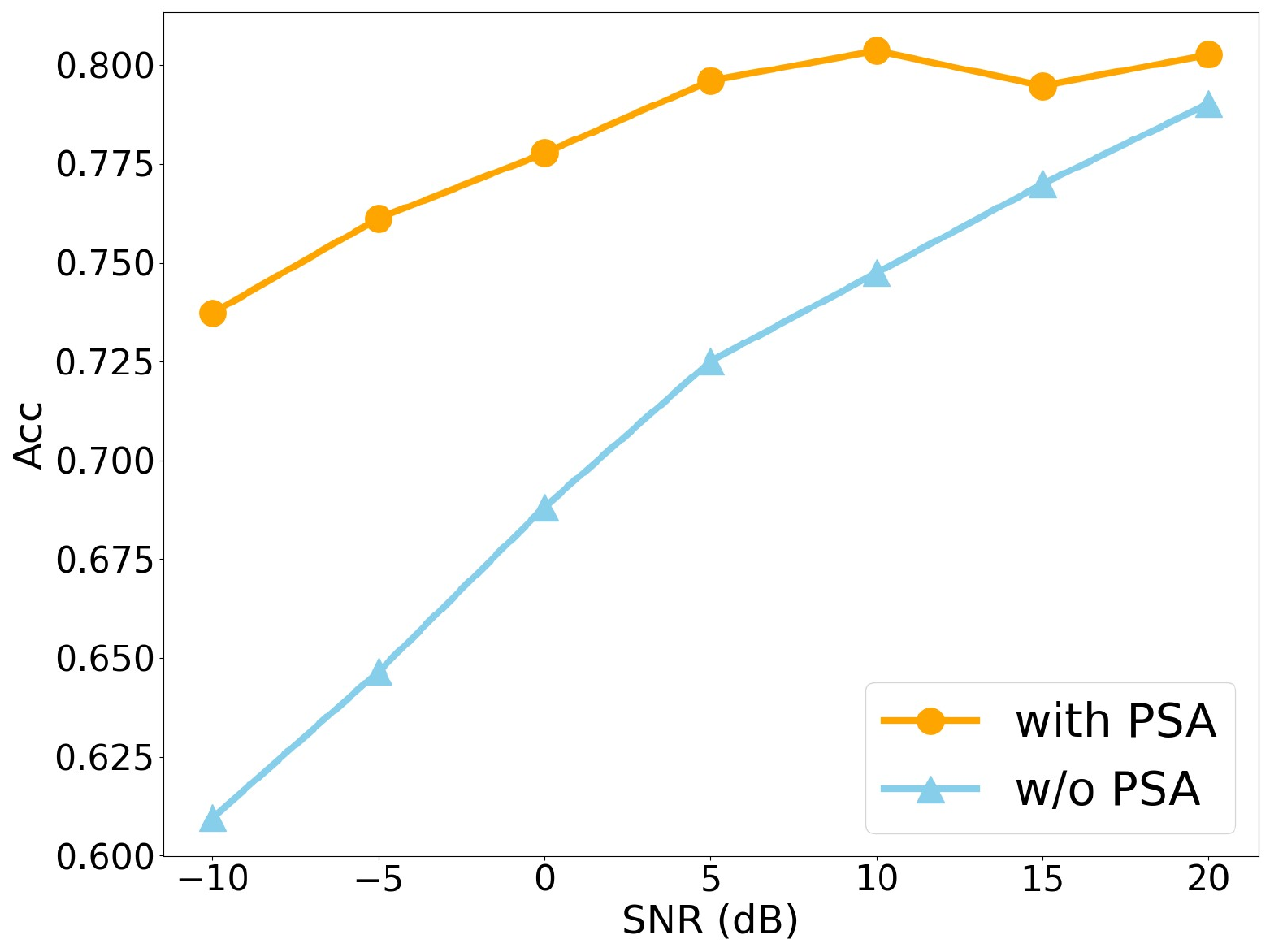}
        \caption*{(b) \ac{Acc}}
    \end{minipage}
    
    \caption{Comparison of model performance with PSA training and without it (i.e., using the pre-trained Whisper backbone) under different SNR levels. 
    }
    \label{fig:abl-PSA}
\end{figure*}
\subsubsection{\ac{PSA} Impact Assessment}

Fig ~\ref{fig:abl-PSA} presents a comparative analysis of \ac{mAP} and \ac{Acc} under varying \ac{SNR} conditions, specifically examining the effects of the \ac{PSA} encoder.
The orange line represents our method, which includes the \ac{PSA} encoder, while the blue line indicates the performance of the model where the \ac{PSA} encoder is replaced by the pre-trained Whisper backbone .
We observe that under strong noise conditions (when the \ac{SNR} is negative), Our method significantly outperforms the model that uses only the audio encoder (baseline configuration).
For instance, at SNR = -10 dB, the model with the \ac{PSA} encoder achieves an \ac{mAP} of approximately 0.72 and an \ac{Acc} of 0.73, whereas the model following baseline configuration achieves an \ac{mAP} of about 0.63 and an \ac{Acc} of 0.65. This demonstrates an improvement of 0.09 in \ac{mAP} and 0.08 in \ac{Acc} when the \ac{PSA}  encoder is used, highlighting its effectiveness in mitigating the impact of noise on performance.
As the \ac{SNR} increases, the gap between the two models narrows, but the \ac{PSA}  encoder still consistently provides better performance across all \ac{SNR} levels, particularly in noisy environments. This analysis underscores the crucial role of the \ac{PSA}  encoder in improving the robustness and \ac{Acc} of the model under challenging conditions.

\subsubsection{\ac{VMMA} Impact Assessment}
% \begin{table}[b]
% \centering
% \caption{The effects of different \ac{VMMA} module settings on the validation set. $\beta$ represents an adaptive parameter trained to optimize the model's performance under the Coarse-Grained setting.}
% \begin{tabular}{c|c|cc}
% \hline
% \textbf{\( p_{\text{vmm}}\) } & \textbf{Threshold} ($\beta_T$) & \textbf{\ac{Acc} (\%)} & \textbf{\ac{mAP} (\%)} \\
% \hline
% None & \ding{55} & 77.40 & 79.37 \\
% \hline
% Fine-Grained & \ding{55} & \textbf{81.66} & 84.53 \\  
% \hline
% \multirow{4}{*}{Coarse-Grained} 
% & $\beta$ & 81.14 & \textbf{84.83} \\  
% & 20\% & 80.29 & 84.07 \\  
% & 50\% & 78.18 & 83.73 \\  
% & 70\% & 78.01 & 83.50 \\  
% \hline
% \end{tabular}
% \label{tab:abl-VMMA}
% \end{table}

Table~\ref{tab:abl-VMMA} comprehensively analyzes the effects of different \ac{VMMA} module settings on the validation set. 
The results indicate that without \ac{VMMA} module, the model achieves an \ac{Acc} of 77.40\% and an \ac{mAP} of 79.37\%. 
When fine-grained \ac{VMMA} is applied, the model's performance significantly improves, reaching 81.66\% \ac{Acc} and 84.53\% \ac{mAP}, demonstrating that detailed and precise modality awareness is crucial for optimal results.
For coarse-grained \ac{VMMA}, the adaptive \(\beta\) parameter yields the best performance, achieving an \ac{Acc} of 81.14\% and a \ac{mAP} of 84.83\%. This suggests that utilizing an adaptive threshold allows for optimal handling of modality loss, as it dynamically adjusts based on the data.
In comparison, when a fixed threshold of 20\% is applied, the performance slightly decreases, with an \ac{Acc} of 80.29\% and an \ac{mAP} of 84.07\%. As the threshold increases to 50\% and 70\%, further declines are observed, with accuracies of 78.18\% and 78.01\%, and \ac{mAP}s of 83.73\% and 83.50\%, respectively. These findings indicate that the adaptive \(\beta\) is more effective for managing missing data in coarse-grained VMMA, while the fine-grained VMMA method continues to provide the highest overall performance, underscoring the significance of detailed modality handling within the multi-modal framework.

\subsection{Qualitative Analysis}

This section provides visual examples of the challenging scenarios addressed by EgoAdapt and compares its performance with other state-of-the-art methods. 
The scenarios include multi-speaker environments, missing visual modalities, and strong background noise, where the model's ability to handle these difficulties is highlighted.

\begin{figure*}[!tb]
\centering
\includegraphics[width=0.98\textwidth]{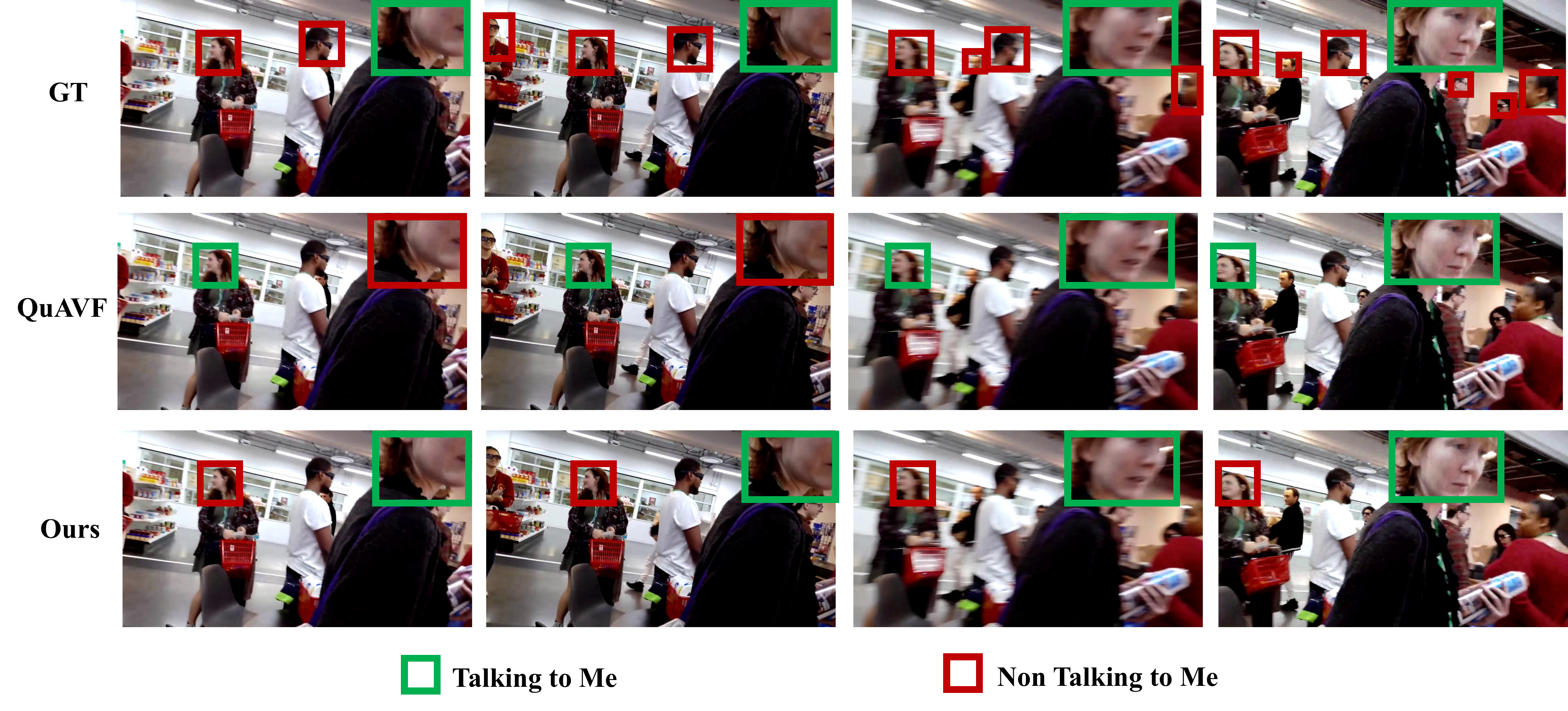}
\caption{Comparison of our proposed method with QuAVF ~\cite{lin2023quavf} (currently the SOTA method) in a multi-speaker environment where multiple individuals are speaking, and the model must correctly identify who is speaking to the camera wearer.}
\label{fig:v-multi_speaker}
\end{figure*}

Figure~\ref{fig:v-multi_speaker} compares our proposed method with QuAVF ~\cite{lin2023quavf}, the current state-of-the-art method, in a multi-speaker environment where multiple individuals speak simultaneously. 
The results reveal that existing methods, which do not account for head orientation, often misidentify other individuals in the scene as addressing the camera wearer. 
In contrast, our method effectively incorporates head orientation, significantly minimizing misidentifications and accurately identifying who directly engages with the camera wearer. 
This distinction is vital in complex environments with multiple speakers, where precise interaction analysis is essential.

\begin{figure}[!tb]
\centering
\includegraphics[width=0.98\textwidth]{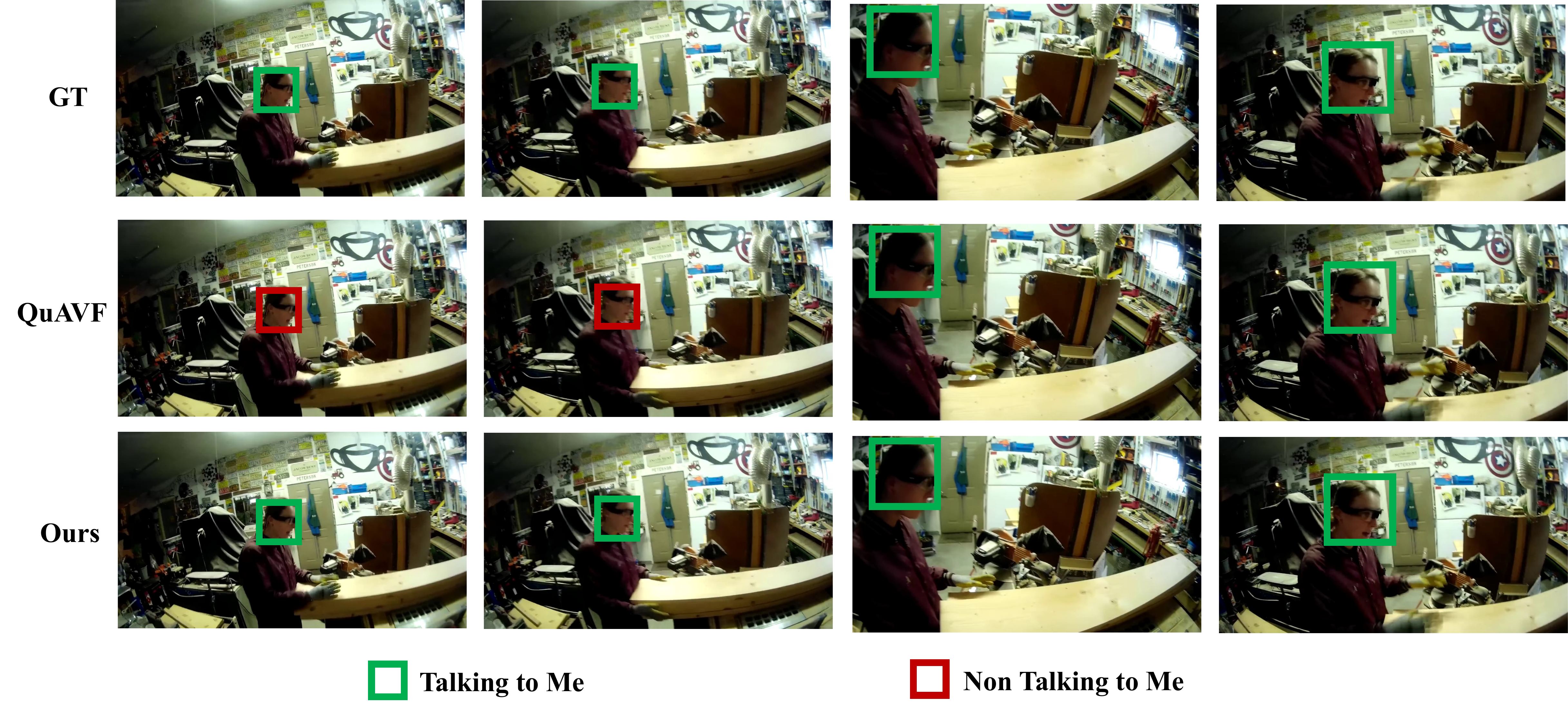}
\caption{Comparison of our proposed method with QuAVF ~\cite{lin2023quavf} in noisy background scenario, where the model must distinguish between relevant and irrelevant audio signals.}
\label{fig:v-noisy_background}
\end{figure}

Figure~\ref{fig:v-noisy_background} illustrates a comparison between the performance of our proposed method and QuAVF in a noisy environment. 
In the first half of the video segment, the speaker is positioned relatively far from the camera wearer, and the noise of the surrounding machinery largely masks the speaker’s voice. 
As a result, the existing method fails to identify the speaker, leading to misclassification correctly. 
In contrast, our method accurately identifies the speaker throughout the video, demonstrating robustness against background noise and maintaining correct speaker identification even when environmental sounds partially obscure the audio signal.

\begin{figure}[!tb]
\centering
\includegraphics[width=0.98\textwidth]{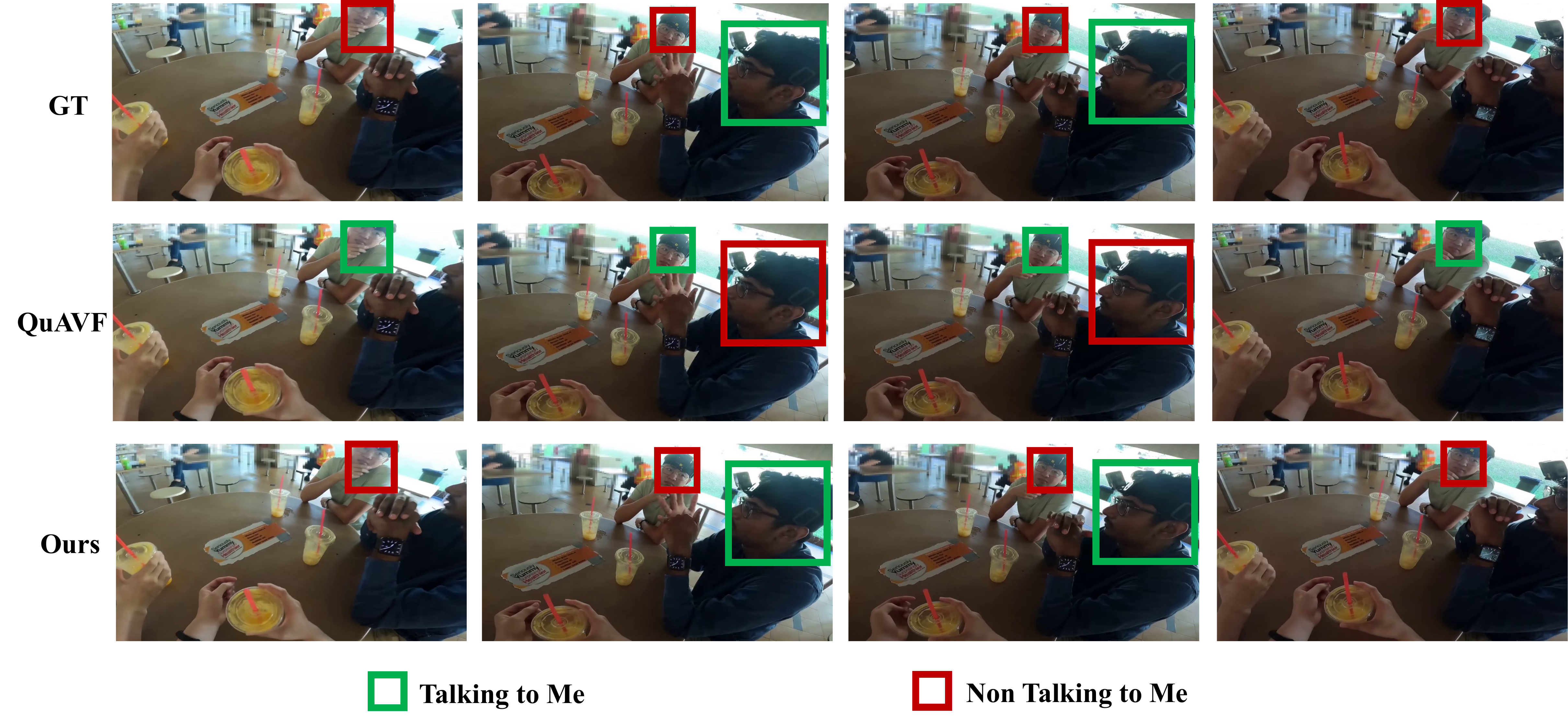}
\caption{Comparison of our proposed method with QuAVF ~\cite{lin2023quavf} in the scenario where the visual modality is partially missing, and the model relies on audio and contextual information to make predictions.}
\label{fig:v-missing_modality}
\end{figure}

Figure~\ref{fig:v-missing_modality} compares the results obtained from our proposed method and QuAVF in scenarios where the visual modality is partially absent. 
Existing methods frequently misidentify a listener—who remains consistently present in the scene—as the individual addressing the camera wearer. 
This misidentification can occur even when the actual speaker is temporarily occluded due to camera movement. 
In contrast, our method effectively distinguishes between the speaker and the listener, accurately identifying the correct speaker even in challenging situations where the speaker may be intermittently out of view. 
This capability is critical for ensuring reliable interaction analysis in dynamic environments.
\subsection{Limitations}
EgoAdapt assumes temporally synchronized visual and auditory inputs, which may limit its applicability in unconstrained real-world recordings where synchronization is imperfect. The multi-stage attention fusion incurs higher computational cost than simpler fusion schemes, making deployment in resource-limited settings more challenging. Moreover, our experiments are conducted on specific egocentric datasets, and the transferability to other domains remains to be explored in future work.

\section{CONCLUSIONS }

In this study, we introduced EgoAdapt, an innovative framework designed to enhance the robustness of egocentric interactive speaker detection in scenarios characterized by missing modalities. 
By leveraging three critical modules—the \acf{VSTR} module, the \acf{PSA} encoder, and the \acf{VMMA} module—EgoAdapt effectively integrates head movements, lip movements, and audio cues. 
The \ac{VSTR} module allows for a nuanced understanding of non-verbal cues and verbal cues, enhancing the model's ability to determine speaker intent accurately. 
Meanwhile, the \ac{PSA} encoder enhances the model's resilience to noisy environments by improving audio feature extraction under challenging conditions. 
The \ac{VMMA} module dynamically assesses the presence or absence of visual information at each frame, allowing EgoAdapt to adapt its reliance on audio or visual cues in real-time. 
Our comprehensive evaluations on the \ac{TTM} benchmark of the Ego4D dataset demonstrate that EgoAdapt not only achieves competitive performance but also surpasses existing baseline methods, affirming its effectiveness in robust interactive speaker detection.

\bibliographystyle{ACM-Reference-Format}
\bibliography{sample-base}

\end{document}

%% file: acro.tex
\newacro{TTM}[TTM]{Talking to Me}
\newacro{MSE}[MSE]{Mean Square Error}
\newacro{ASR}[ASR]{automatic speech recognition}
\newacro{VMMA}[VMMA]{Visual Modality Missing Awareness}
\newacro{mAP}[mAP]{mean average precision}
\newacro{HPE}[HPE]{Head Pose Estimation}
\newacro{VSTR}[VSTR]{Visual Speaker Target Recognition}
\newacro{PSA}[PSA]{Parallel Shared-weight Audio} 
\newacro{SNR}[SNR]{signal-to-noise ratios}
\newacro{Acc}[Acc]{accuracy}
\newacro{VIT}[VIT]{Vision Transformer}